\documentclass[aps,prb,twocolumn,showpacs,groupedaddress]{revtex4}

\usepackage{graphicx}
\usepackage{dcolumn}
\usepackage{bm}
\usepackage{amssymb}

\begin{document}


\title{ 
Solvent-free coarse-grained lipid model for large-scale simulations}

\author{Hiroshi Noguchi}
\email[]{noguchi@issp.u-tokyo.ac.jp}
\affiliation{
Institute for Solid State Physics, University of Tokyo,
 Kashiwa, Chiba 277-8581, Japan}

\date{\today}

\begin{abstract}
A coarse-grained molecular model, which consists of a spherical particle
and an orientation vector, is proposed to simulate lipid membrane on a large length scale.
The solvent is implicitly represented by an effective attractive interaction between particles.
A bilayer structure is formed by orientation-dependent (tilt and bending) potentials.
In this model, the membrane properties (bending rigidity, line tension of membrane edge,
area compression modulus, lateral diffusion coefficient, and flip-flop rate) can be varied over broad ranges.
The stability of the bilayer membrane is investigated via droplet-vesicle transition.
The rupture of the bilayer and worm-like micelle formation can be induced by an increase in the spontaneous curvature of the monolayer membrane.

\end{abstract}


\maketitle

\section{Introduction}

Amphiphilic molecules, such as lipids and detergents,
 self-assemble into various structures depending on the relative
size of their hydrophilic parts:
spherical or worm-like micelles, bilayer membranes,
inverted hexagonal structures, and inverted micelles.
Among these structures, the bilayer membrane of phospholipids 
has been intensively investigated, since it is the basic structure
of the plasma membrane and the intracellular compartments of living cells,
where the membranes are in a fluid phase and lipid molecules can diffuse 
in quasi-two-dimensional space.
A vesicle (closed membrane) is considered to be a simple model of cells
and it has applications in drug-delivery systems as a drug carrier.

Bilayer membranes exhibit many interesting phenomena such as shape deformation induced by phase separation or
chemical reaction,  membrane fusion, and membrane fission.
The length scale of these phenomena varies from nm to $\mu$m,
since cells are $\sim 10\mu$m in diameter, 
whereas the thickness of a biomembrane is $5$nm.
To investigate the morphologies of cells and vesicles,
the molecular structure is assumed to be negligible, and
the bilayer membrane is described as a smoothly-curved mathematical 
surface \cite{safr94,lipo95,seif97}.
The information about the bilayer properties is only 
reflected in the values of the elastic parameters.
To simulate the membrane with thermal fluctuations, a triangulated surface is widely used \cite{gg:gomp04c,gg:gomp97f}.
An alternative model is a meshless membrane \cite{drou91,nogu06,nogu06a,popo08,kohy09,liu09,fuch09,yuan10},
where particles self-assemble into a membrane by anisotripic potential interactions.
These models can reproduce $\mu$m-scale dynamics of the bilayer membrane well
but cannot treat a non-bilayer structure such as the stalk structure of a membrane fusion intermediate \cite{jahn02,cher08}.

To simulate molecular-scale dynamics and the non-bilayer structure,
a molecular model is required.
Although computer technology has grown rapidly,
the typical scale for recent simulations of the all-atom models
is only $100$ ns dynamics of hundreds of lipid molecules.
To simulate the membranes on longer and larger scales,
various coarse-grained molecular models have been proposed (see review articles \cite{muel06,vent06,klie08,nogu09,marr09}).
Recently, the potential parameters in some of the coarse-grained molecular models are tuned 
by atomistic simulations \cite{marr04,izve05,arkh08,shin08,wang10}.
In mapping of interaction parameters, one coarse-grained particle typically 
represents three or four heavy atoms and their accompanying hydrogen atoms.
To further reduce the computational costs,
larger segments (three or more segment particles per amphiphilic molecule)
are employed, and the solvent is implicitly represented by
an effective attractive potential between the hydrophobic segments \cite{nogu09,nogu01a,nogu03,bran06,dese09,fara09}.
Model parameters are chosen to generate a bilayer membrane with reasonably realistic values of elastic properties.
In this scale, it is difficult to take into account chemical details of lipids, such as an unsaturated bond in hydrocarbon chains.
Instead, this type of models can be advantageous to capture the general features in the bilayer membrane,
since the simplicity of the model can allow for wide ranges of variation of the membrane  properties.

In this paper, we propose a solvent-free molecular model to pursue two purposes:
(1) to represent the amphiphilic molecule in a size as small as possible and
(2) to allow the variation in the membrane properties for wide ranges. 
A molecule consisting of many particles
has a higher resolution than that with less particles
but requires a smaller length unit and time step for simulations.
Here, we consider a molecule that consists of a spherical particle and an orientation vector.
It can reduce computational costs to simulate many molecules.
In previous solvent-free models \cite{nogu09,wang10,nogu01a,nogu03,bran06,dese09,fara09},
the  membrane properties are varied only in narrow ranges.
On the other hand, in one of the meshless membrane models, 
the bending rigidity and the line tension of the membrane edge
can be independently varied over wide ranges \cite{nogu06}.
This allows the conditions of vesicle formation and rupture to be controlled \cite{nogu06a}.
In addition, it is easy to compare the simulation results with theoretical predictions.
Such tuning capability is desired for molecular models.

In Sec. \ref{sec:method}, the lipid model and the simulation method are described.
In Sec. \ref{sec:results}, the results and discussion are provided.
The formation of a membrane and its stability for droplet-vesicle transitions
are described in Sec. \ref{sec:selfassem}. 
In Sec. \ref{sec:mempro} and \ref{sec:paradep}, the calculation methods
and the dependence of the static and dynamic properties on model parameters are described, respectively.
The summary is given in Sec. \ref{sec:sum}.

\section{Model and Method}
\label{sec:method}

\subsection{Molecular model}

\begin{figure}
\includegraphics{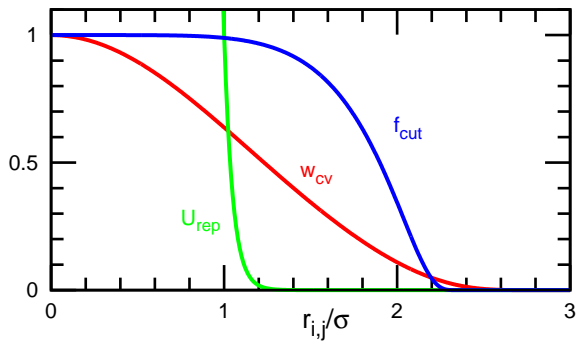}
\caption{\label{fig:pot}
(Color online)
The cutoff function $f_{\rm {cut}}(r_{ij})$,
the compact Gaussian weight function $w_{\rm {cv}}(r_{ij})$, and
the repulsive potential $U_{\rm {rep}}(r_{ij})$.
}
\end{figure}

In solvent-free lipid models, an amphiphilic molecule is typically represented by three or more particles \cite{nogu09}.
Here, we consider a lipid molecule with minimum ($5$) degrees of freedom for solvent-free molecular simulations.
Each ($i$-th) molecule has a spherical particle with an orientation vector ${\bf u}_i$, 
which represents a direction from the hydrophobic to the hydrophilic part.
There are two points of interaction in the molecule:
the center of a sphere ${\bf r}^{\rm s}_i$ and a hydrophilic point ${\bf r}^{\rm e}_i={\bf r}^{\rm s}_i+{\bf u}_i$.
The molecules interact with each other via the potential,
\begin{eqnarray}
\frac{U}{k_{\rm B}T} &=\ \ & \hspace{1cm} \sum_{i<j} U_{\rm {rep}}(r_{i,j}^{\rm s}) \label{eq:U_all}
               +\varepsilon \sum_{i} U_{\rm {att}}(\rho_i)  \\ \nonumber
&\ \ +& \ \ \frac{k_{\rm{tilt}}}{2} \sum_{i<j} \bigg[ 
( {\bf u}_{i}\cdot \hat{\bf r}^{\rm s}_{ij})^2
 + ({\bf u}_{j}\cdot \hat{\bf r}^{\rm s}_{ij})^2  \bigg] w_{\rm {cv}}(r^{\rm e}_{ij}) \\ \nonumber
&\ \ +&  \frac{k_{\rm {bend}}}{2} \sum_{i<j}  \bigg({\bf u}_{i} - {\bf u}_{j} - C_{\rm {bd}} \hat{\bf r}^{\rm s}_{ij} \bigg)^2 w_{\rm {cv}}(r^{\rm e}_{ij}),
\end{eqnarray} 
where ${\bf r}_{i,j}={\bf r}_{i}-{\bf r}_j$, $r_{i,j}=|{\bf r}_{i,j}|$,
 $\hat{\bf r}_{i,j}={\bf r}_{i,j}/r_{i,j}$, and $k_{\rm B}T$ is the thermal energy.
The molecules have an excluded volume with a diameter $\sigma$ via the repulsive potential,
\begin{equation}
\label{eq:rep}
U_{\rm {rep}}(r)=\exp[-20(r/\sigma-1)],
\end{equation}
with a cutoff at $r=2.4\sigma$.

The second term in Eq. (\ref{eq:U_all}) represents the attractive interaction between the molecules.
A multibody attractive potential $U_{\rm {att}}(\rho_i)$ is 
employed to mimic the ``hydrophobic'' interaction.
This potential allows the formation of  the fluid membrane over wide parameter ranges
and fast lateral diffusion.
Similar potentials have been applied in the previous membrane models \cite{nogu01a,nogu06,fara09}
 and a coarse-grained protein model \cite{taka99}.
The potential $U_{\rm {att}}(\rho_i)$ is given by
\begin{eqnarray} \label{eq:U_att}
U_{\rm {att}}(\rho_i) = 0.25\ln[1+\exp\{-4(\rho_i-\rho^*)\}]- C,
\end{eqnarray} 
where $C= 0.25\ln\{1+\exp(4\rho^*)\} \simeq \rho^*$ is chosen such that 
$U_{\rm {att}}(0)=0$.
The local particle density $\rho_i$ is approximately the number of
particles ${\bf r}^{\rm s}_i$  in the sphere whose radius is $r_{\rm {att}}$.
\begin{equation}
\rho_i= \sum_{j \ne i} f_{\rm {cut}}(r^{\rm s}_{i,j}), 
\label{eq:wrho}
\end{equation} 
where $f_{\rm {cut}}(r)$ is a $C^{\infty}$ cutoff function,
\begin{equation} \label{eq:cutoff}
f_{\rm {cut}}(r)=\left\{ 
\begin{array}{ll}
\exp\{A(1+\frac{1}{(r/r_{\rm {cut}})^n -1})\}
& (r < r_{\rm {cut}}) \\
0  & (r \ge r_{\rm {cut}}) 
\end{array}
\right.
\end{equation}
with $n=6$, $A=\ln(2) \{(r_{\rm {cut}}/r_{\rm {att}})^n-1\}$,
$r_{\rm {att}}= 1.9\sigma$  $(f_{\rm {cut}}(r_{\rm {att}})=0.5)$, 
and the cutoff radius $r_{\rm {cut}}=2.4\sigma$ (see Fig. \ref{fig:pot}). 
The potential $U_{\rm {att}}(\rho_i)$ acts as a pairwise attractive potential 
($U_{\rm {att}}(\rho_i)\simeq \rho$, so that $\sum_i U_{\rm {att}}(\rho_i)\simeq - 2\sum_{i<j} f_{\rm {cut}}(r^{\rm s}_{i,j})$) 
for $\rho_i < \rho^*-1$
and approaches a constant value ($U_{\rm {att}}(\rho_i)\simeq \rho^*$) for $\rho_i > \rho^*+1$.
It is assumed that the hydrophobic parts have no contact with the implicit solvent (void space) at $\rho_i \gtrsim \rho^*$.

\begin{figure}
\includegraphics{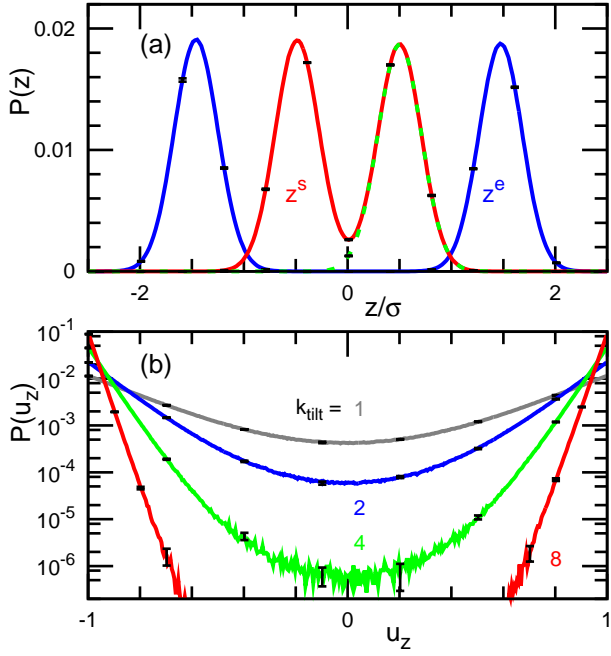}
\caption{\label{fig:hisze}
(Color online)
Probability distribution of (a) the positions and (b) the orientation of
the molecules in the planar membrane at $N=512$,
 $\rho^*=14$, $\varepsilon=2$, $k_{\rm {bend}}=k_{\rm {tilt}}$, and $C_{\rm {bd}}=0$.
(a) The $z$ components of  ${\bf r}^{\rm s}_i$ and ${\bf r}^{\rm e}_i$ are shown for
 $k_{\rm {tilt}}=8$.
The dashed line represents  ${\bf r}^{\rm s}_i$ of the molecules at $u_z>0$.
(b) The $z$ component $u_z$ of the molecular orientation is shown
for $k_{\rm {tilt}}=1$, $2$, $4$, and $8$.
The error bars are displayed at several data points.
}
\end{figure}

The third and fourth terms in Eq.~(\ref{eq:U_all}) are
discretized versions of
tilt and bending potentials of the tilt model \cite{hamm98,hamm00}, respectively.
A smoothly truncated Gaussian function~\cite{nogu06} 
is employed as a weight function 
\begin{equation} \label{eq:wcv}
w_{\rm {cv}}(r)=\left\{ 
\begin{array}{ll}
\exp (\frac{(r/r_{\rm {ga}})^2}{(r/r_{\rm {cc}})^n -1})
& (r < r_{\rm {cc}}) \\
0  & (r \ge r_{\rm {cc}}) 
\end{array}
\right.
\end{equation}
with  $n=4$, $r_{\rm {ga}}=1.5\sigma$, and $r_{\rm {cc}}=3\sigma$ (see Fig. \ref{fig:pot}). 
All orders of derivatives of $f_{\rm {cut}}(r)$ and $w_{\rm {mls}}(r)$ 
are continuous at the cutoff radii.
The weight is a function of $r^{\rm e}_{ij}$ (not $r^{\rm s}_{ij}$)
to avoid the interaction between the molecules in the opposite monolayers of the bilayer.
The average distance between the neighboring molecules
in the same monolayer is $\bar{r}_{\rm {nb}} \simeq 1.05\sigma$,
and the distance to the neighboring molecule in the other monolayer 
is $r^{\rm e}_{ij} \simeq 3\sigma$ (see Fig. \ref{fig:hisze}(a)).
Thus, these two potentials act between the neighboring molecules
in the same monolayer but not between the monolayers.
The tilt potential has the energy minimum in a completely flat membrane with no tilt deformation.
Similar tilt potentials have been used in the meshless membrane models \cite{drou91,popo08,kohy09,liu09,fuch09,yuan10}.
The same type of bending potential [the fourth term in Eq. (\ref{eq:U_all})] was
previously used to control the bending rigidity and the spontaneous curvature of the monolayer in the molecular simulations \cite{nogu03,fara09}.
Positive spontaneous curvature indicates that the hydrophilic head is larger than the hydrophobic tail of amphiphilic molecules.
The bending rigidity is numerically calculated and compared with the 
estimation from the continuous description of the membrane in Sec. \ref{sec:paradep}.

\subsection{Brownian dynamics}

We simulated the membrane in the $NVT$ ensemble (constant
number of molecules $N$, volume $V$, and temperature $T$) with periodic boundary 
conditions in a box with side length $L_x$, $L_y$, and $L_z$.
We employed Brownian dynamics (molecular dynamics with Langevin thermostat).
The motions of the center of the mass 
${\bf r}^{\rm G}_{i}=({\bf r}^{\rm s}_{i}+{\bf r}^{\rm e}_{i})/2 $ and 
the orientation ${\bf u}_{i}$ are given by underdamped Langevin equations:
\begin{eqnarray}
  \frac{d {\bf r}^{\rm G}_{i}}{dt} &=& {\bf v}^{\rm G}_{i}, \ \  \frac{d {\bf u}_{i}}{dt} = {\boldsymbol \omega}_{i}, \\
m \frac{d {\bf v}^{\rm G}_{i}}{dt} &=&
 - \zeta_{\rm G} {\bf v}^{\rm G}_{i} + {\bf g}^{\rm G}_{i}(t)
 + {\bf f}^{\rm G}_i, \\
I \frac{d {\boldsymbol \omega}_{i}}{dt} &=&
 - \zeta_{\rm r} {\boldsymbol \omega}_i + ({\bf g}^{\rm r}_{i}(t)
 + {\bf f}^{\rm r}_i)^{\perp} + \lambda {\bf u}_{i},
\end{eqnarray}
where $m$ and $I$ are the mass and the moment of inertia of the molecule, respectively.
The forces are given by ${\bf f}^{\rm G}_i= - \partial U/\partial {\bf r}^{\rm G}_{i}$
and ${\bf f}^{\rm r}_i= - \partial U/\partial {\bf u}_{i}$ with the perpendicular component ${\bf a}^{\perp} ={\bf a}- ({\bf a}\cdot{\bf u}_{i}) {\bf u}_{i}$
and a Lagrange multiplier $\lambda$ to keep ${\bf u}_{i}^2=1$.
According to  the fluctuation-dissipation theorem,
the friction coefficients $\zeta_{\rm G}$, $\zeta_{\rm r}$ and 
the Gaussian white noises ${\bf g}^{\rm G}_{i}(t)$,  ${\bf g}^{\rm r}_{i}(t)$
obey the following relations:
the average $\langle g^{\beta_1}_{i,\alpha_1}(t) \rangle  = 0$ and the variance
$\langle g^{\beta_1}_{i,\alpha_1}(t) g^{\beta_2}_{j,\alpha_2}(t')\rangle  =  
         2 k_{\rm B}T \zeta_{\beta_1} \delta _{ij} \delta _{\alpha_1 \alpha_2} \delta _{\beta_1 \beta_2} \delta(t-t')$,
where $\alpha_1, \alpha_2 \in \{x,y,z\}$ and  $\beta_1, \beta_2 \in \{{\rm G, r}\}$.
The Langevin equations are integrated by the leapfrog algorithm \cite{alle87} with ${\bf v}_{i,n} \equiv {\bf v}_{i}(t_{n}) = ({\bf v}_{i}(t_{n+1/2})+{\bf v}_{i}(t_{n-1/2}))/2$.
First, the velocities are updated by
\begin{eqnarray} \label{eq:lp}
{\bf v}^{\rm G}_{i,{n+1/2}} &=& a_0{\bf v}^{\rm G}_{i,{n-1/2}} + a_1({\bf g}^{\rm G}_{i,n} + {\bf f}^{\rm G}_{i,n}),\\ \nonumber
{\boldsymbol \omega}'_{i,{n+1/2}} &=& b_0{\boldsymbol \omega}_{i,{n-1/2}} + b_1({\bf g}^{\rm r}_{i,n} + {\bf f}^{\rm r}_{i,n})^{\perp} + \lambda' {\bf u}_{i,n},\\ \nonumber
{\bf u}'_{i,{n+1/2}} &=& {\bf u}_{i,{n}}+{\boldsymbol \omega}'_{i,{n+1/2}}\Delta t/2, \\ \nonumber
{\boldsymbol \omega}_{i,{n+1/2}} &=& {\boldsymbol \omega}'_{i,{n+1/2}} -  ({\boldsymbol \omega}'_{i,{n+1/2}}\cdot {\bf u}'_{i,{n+1/2}}){\bf u}'_{i,{n+1/2}},
\end{eqnarray}
where
\begin{eqnarray}
 a_0 &=& \frac{1-\zeta_{\rm G} \Delta t/2m}{1+\zeta_{\rm G} \Delta t/2m},\  
a_1= \frac{\Delta t/m}{1+\zeta_{\rm G} \Delta t/2m}, \\  \nonumber
b_0 &=& \frac{1-\zeta_{\rm r} \Delta t/2I}{1+\zeta_{\rm r} \Delta t/2I}, \  \ 
b_1= \frac{\Delta t/I}{1+\zeta_{\rm r} \Delta t/2I},  \\ \nonumber
\lambda' &=& \lambda \Delta t/I = -\frac{2{\boldsymbol \omega}_{i,{n-1/2}}\cdot{\bf u}_{i,n}}{1+\zeta_{\rm r} \Delta t/2I}, \\ \nonumber
{\bf g}^{\beta_1}_{i,n} &=& {\bf g}^{\beta_1}_i(t_n)/\sqrt{\Delta t}.
\end{eqnarray}
Then, the positions are updated by
\begin{eqnarray}
 {\bf r}^{\rm G}_{i,{n+1}} &=& {\bf r}^{\rm G}_{i,{n}} + {\bf v}^{\rm G}_{i,{n+1/2}}\Delta t, \\ \nonumber
 {\bf u}'_{i,{n+1}} &=& {\bf u}_{i,{n}} + {\boldsymbol \omega}_{i,{n+1/2}}\Delta t, \\ \nonumber
 {\bf u}_{i,{n+1}} &=& {\bf u}'_{i,{n+1}}/|{\bf u}'_{i,{n+1}}|. 
\end{eqnarray}

We employed $m=1$, $I=1$, $\zeta_{\rm G}=1$, $\zeta_{\rm r}=1$, $k_{\rm B}T=1$,
 $\Delta t=0.005$, and the total number of the molecules $N=300$ to $8192$.
The results are displayed with the length unit $\sigma$, the energy unit $k_{\rm B}T$, and
 the time unit $\tau_0=\zeta_{\rm G}\sigma^2/k_{\rm B}T$.
The diffusion coefficient $D$ is normalized using the diffusion coefficient $D_0=\sigma^2/\tau_0$ of an isolated molecule.
The error bars of the data are estimated 
from the standard deviations of three to six independent runs.

\begin{figure}
\includegraphics{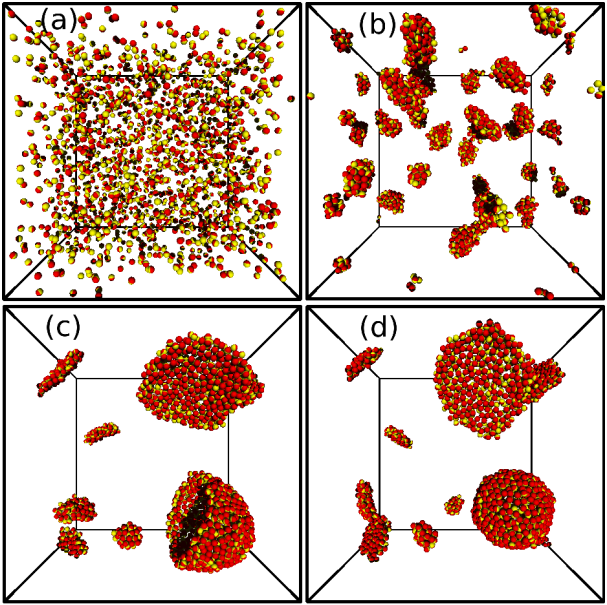}
\caption{\label{fig:ass}
(Color online)
Sequential snapshots of the molecular self-assembly
at $N=2000$, $L_{x}=L_{y}=L_{z}=40\sigma$, $\varepsilon=2$, $\rho^*=14$, 
$k_{\rm {bend}}=0$, $k_{\rm {tilt}}=8$, and $C_{\rm {bd}}=0$.
(a) $t/\tau_0=0$. (b) $t/\tau_0=500$. (c) $t/\tau_0=16650$. (d) $t/\tau_0=16700$. 
}
\end{figure}

\begin{figure}
\includegraphics{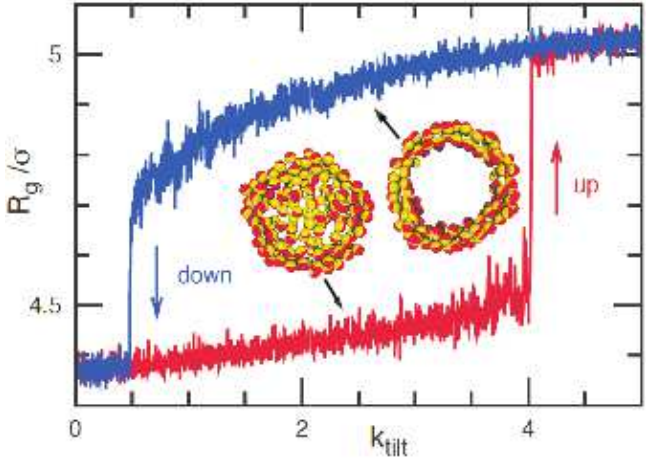}
\caption{\label{fig:fsp_rg}
(Color online)
Droplet-vesicle transition
at $N=500$, $\varepsilon=2$, $\rho^*=14$, $k_{\rm {bend}}=4$, and $C_{\rm {bd}}=0$.
The lower and upper lines represent the radius of gyration $R_{\rm g}$ 
in $k_{\rm {tilt}}$ increasing or decreasing, respectively.
Sliced snapshots are also shown at $k_{\rm {tilt}}=2.5$.
}
\end{figure}

\begin{figure}
\includegraphics[width=8cm]{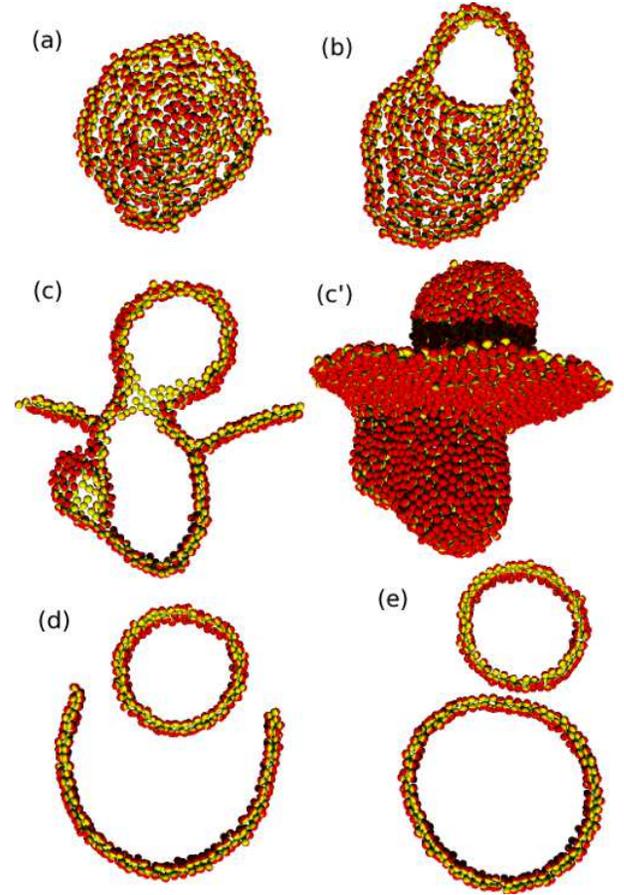}
\caption{\label{fig:ves_snap}
(Color online)
Formation of vesicles from a droplet
at $N=4000$, $\varepsilon=2$, $\rho^*=14$, $k_{\rm {bend}}=8$, and $C_{\rm {bd}}=0$.
The tilt coefficient $k_{\rm {tilt}}$ is gradually increased as $k_{\rm {tilt}}= 0.0005 t/\tau_0$.
Sliced snapshots are shown at
(a) $t/\tau_0=12500$ ($k_{\rm {tilt}}=6.25$), (b) $t/\tau_0=13600$, 
(c) $t/\tau_0=13700$, (d) $t/\tau_0=14000$, and  (e) $t/\tau_0=14500$ ($k_{\rm {tilt}}=7.25$). 
All molecules are also shown for $t/\tau_0=13700$ in (c').
}
\end{figure}

\section{Results and Discussion}
\label{sec:results}

\subsection{Self-assembly and membrane stability}
\label{sec:selfassem}

Molecules self-assemble into spherical droplets at $k_{\rm {bend}}=k_{\rm {tilt}}=0$, i.e., 
when only the first two terms in Eq. (\ref{eq:U_all}) are taken into account.
When the third term (the tilt potential) is added,
the molecules can spontaneously form vesicles.
Figure \ref{fig:ass} shows the self-assembly of molecules from a random gas state.
First, small clusters are formed; these clusters merge into disk-like micelles.
Then, a large disk closes into a vesicle through a bowl-like shape (see Fig. \ref{fig:ass}(c)).
Similar self-assembly processes have been observed in the previous simulations of molecular~\cite{nogu01a} 
and meshless~\cite{nogu06a} models.

In order to clarify the stability of three-dimensional aggregates and bilayer membranes, the morphologies of the aggregates are investigated as $k_{\rm {tilt}}$ gradually increases or decreases.
As $k_{\rm {tilt}}$ increases, a spherical liquid droplet transforms into a bilayer vesicle. At the transition point ($k_{\rm {tilt}}=4$), the radius of gyration $R_{\rm g}$
exhibits an abrupt increase as shown in Fig. \ref{fig:fsp_rg}.
As $k_{\rm {tilt}}$ decreases, the transition from a vesicle to a droplet occurs, however,
the transition point ($k_{\rm {tilt}}=0.5$) is much lower.
Thus, a typical hysteresis for the first-order transition is observed. 
The rate of increase or decrease is sufficiently low ($k_{\rm {tilt}}= 0.0005 t/\tau_0$). 
We checked that the deviation of the transition points by the annealing rates is very small; $\Delta k_{\rm {tilt}}=0.1$ between $k_{\rm {tilt}}= 0.000125 t/\tau_0$ and $k_{\rm {tilt}}= 0.001 t/\tau_0$ 
with  $N=500$, $\varepsilon=2$, $\rho^*=14$, $k_{\rm {bend}}=k_{\rm {tilt}}$, and $C_{\rm {bd}}=0$.
The transition points are not sensitive to the path of $k_{\rm {bend}}(k_{\rm {tilt}})$, since the difference of the results for $k_{\rm {bend}}=k_{\rm {tilt}}$ and constant $k_{\rm {bend}}$ is smaller than their statistical errors.

\begin{figure}
\includegraphics{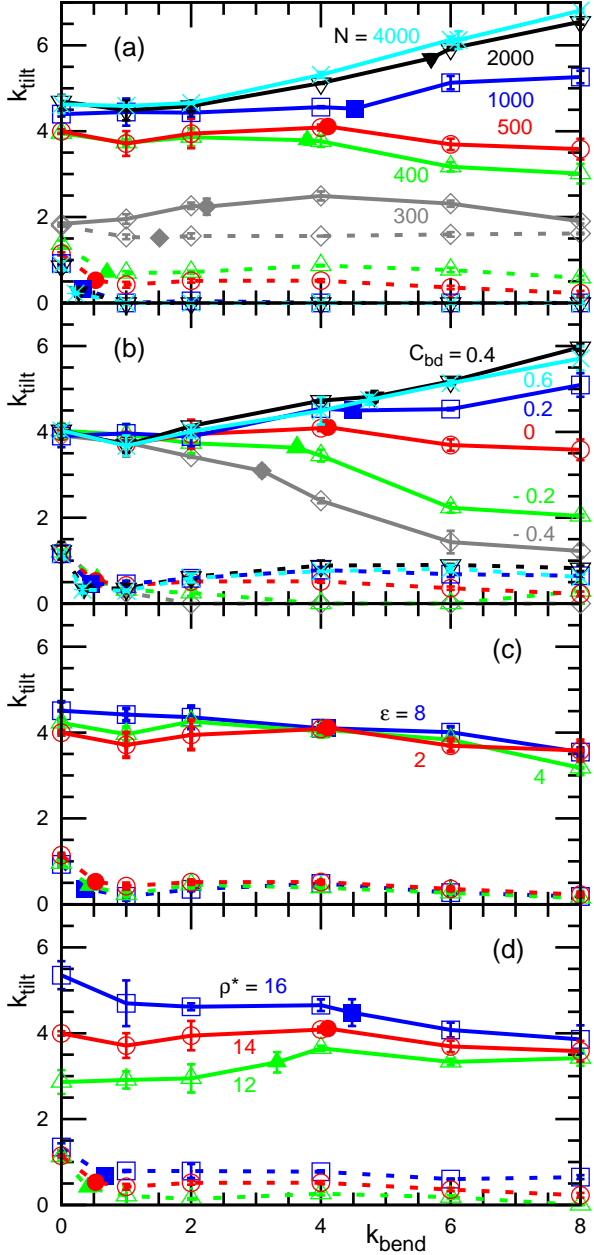}
\caption{\label{fig:fsp_nc}
(Color online)
Shape transition points of molecular aggregates
between the droplet and the bilayer membrane (vesicles and disks)
for various (a) $N$, (b) $C_{\rm {bd}}$, (c) $\varepsilon$, and (d) $\rho^*$.
If not specified, $N=500$, $\varepsilon=2$, $\rho^*=14$, and $C_{\rm {bd}}=0$.
The solid and dashed lines represent data
with increasing and decreasing $k_{\rm {tilt}}$, respectively.
The open and filled symbols represent data
with fixed $k_{\rm {bend}}$ and $k_{\rm {bend}}=k_{\rm {tilt}}$, respectively.
}
\end{figure}

\begin{figure}
\includegraphics{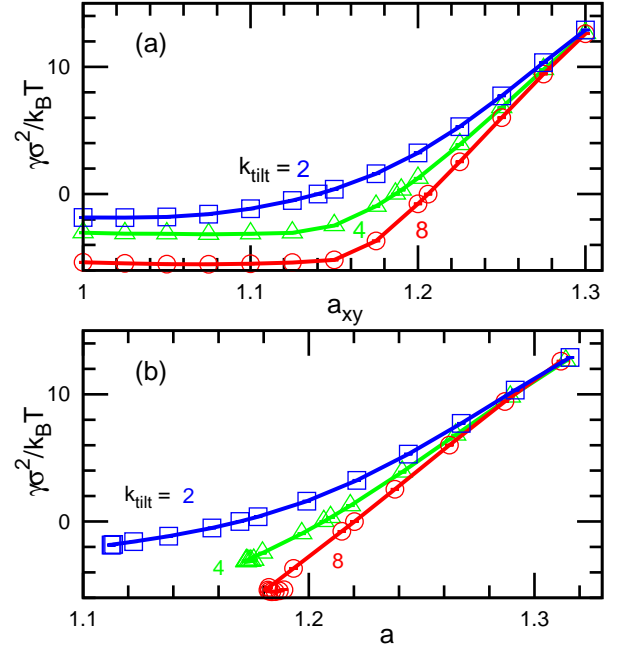}
\caption{\label{fig:ten}
(Color online)
Surface tension $\gamma$ of a flat membrane
at $N=512$, $\rho^*=14$, $\varepsilon=2$,
$k_{\rm {bend}}=k_{\rm {tilt}}$, and $C_{\rm {bd}}=0$.
Dependence of $\gamma$ on (a) the projected area per molecule $a_{xy}=2A_{xy}/N\sigma^2$
and (b) the intrinsic area per molecule $a= 2A/N\sigma^2$.
The squares, triangles, and circles represent $\gamma$ for $k_{\rm {tilt}}=2, 4$, and $8$,
respectively. The error bars are smaller than the line thickness.
}
\end{figure}

For large aggregates with $N=4000$, two vesicles or a vesicle with disks
are formed instead of a single vesicle.
Figure \ref{fig:ves_snap} shows an example of the formation of two vesicles.
A void space is opened in the droplet, and a bilayer skirt is formed.
Then, it is separated into two parts and forms two vesicles.
If the separated membrane is small, a disk is formed.
Since the larger droplets can have a clearer molecular layer on the surface,
which prevents the shape change,
higher $k_{\rm {tilt}}$ is needed to trigger the shape transition [see Fig. \ref{fig:fsp_nc}(a)].
At $N=300$, the coexistence region of the droplets and the vesicles is narrow,
since the number of the molecules is not sufficient to form the surface and inside layers. 
The points of the droplet-bilayer transition are also dependent on $C_{\rm {bd}}$,
while they are almost independent of $\varepsilon$ and $\rho^*$.
Thus, the bilayer stability is determined by the tilt and bending potentials
but not by the attractive potential.

Let us discuss the condition required to form a stable bilayer.
When all molecules have the same orientation ${\bf u}_i={\bf u}_j$,
 the bending potential energy becomes zero at $C_{\rm {bd}}=0$.
Thus, the bending potential with $C_{\rm {bd}}=0$ 
can have the minimum energy for any structure of the aggregate.
Therefore, the spherical liquid droplet, which has the minimum surface area, 
would be the equilibrium state at $k_{\rm {tilt}}=0$ instead of the bilayer.
However, large vesicles with $N \ge 1000$ and planar membranes 
can maintain their bilayer structure as a metastable state
even at $k_{\rm {tilt}}=0$ with finite $k_{\rm {bend}}$.
Note that the capability to keep a pre-formed bilayer membrane
 does not guarantee the self-assembly to the bilayer.
In particular, the periodic boundary condition 
is a strong constraint, which can 
keep the bilayer membrane as a thin liquid layer even at $k_{\rm {tilt}}=k_{\rm {bend}}=0$.
In order to obtain the spontaneous formation of the bilayer membrane,
$k_{\rm {tilt}} > 2$ is required.

\subsection{Calculation of membrane properties}
\label{sec:mempro}

To investigate the membrane properties, we formed a nearly planar membrane
 without edges or pores.
The membrane area and the surface tension are varied by increasing or decreasing 
the projected area $A_{xy}=L_x L_y$, where $L_x=L_y$.
The membrane has a clear bilayer structure (see Fig. \ref{fig:hisze}).
The intrinsic area of the tensionless membrane per molecule $a_0=2A_0/N\sigma^2$, 
the area compression modulus $K_{\rm A}$,
and the half lifetime $\tau_{\rm {ff}}$ of the flip-flop motion are calculated 
from the flat membranes with $N=512$.
The bending rigidity $\kappa$ and
the diffusion coefficient $D$ are calculated at larger tensionless membranes
with $N=8192$.
The line tension $\Gamma$ of membrane edge  is calculated 
from the strip of the flat membrane with $N=512$.

\begin{figure}
\includegraphics{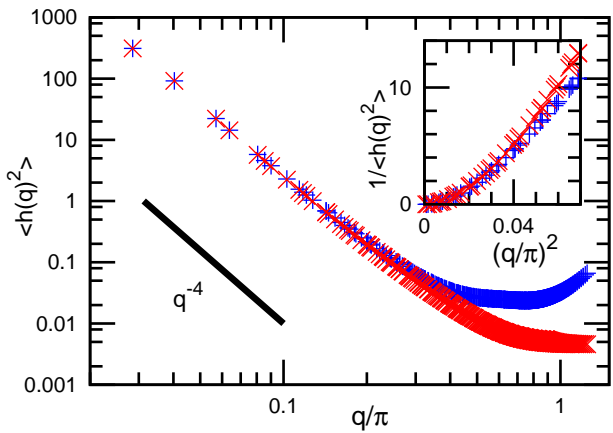}
\caption{\label{fig:hq}
(Color online)
Spectra of undulation modes $\langle |h(q)|^2 \rangle$ of nearly planar, 
tensionless membranes ($\gamma=0$)
at  $N=8192$, $\rho^*=14$, $\varepsilon=2$,
$k_{\rm {bend}}=k_{\rm {tilt}}=8$, and $C_{\rm {bd}}=0$.
Results for $\langle |h(q)|^2 \rangle$ calculated from the molecular positions
($+$) and from the averaged positions on a square mesh ($\times$) are shown.
The inset shows the dependence of $1/\langle |h(q)|^2 \rangle$ on $q^2$, 
which is used to extract the bending rigidity $\kappa$. 
}
\end{figure}

\begin{figure}
\includegraphics{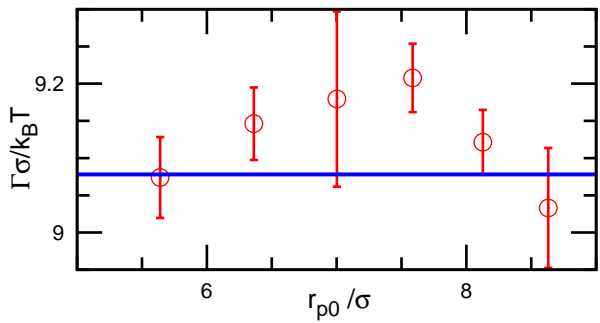}
\caption{\label{fig:line_r}
(Color online)
Line tension $\Gamma$ of membrane edge
at $\varepsilon=2$, $\rho^*=14$, $k_{\rm {bend}}=4$, $k_{\rm {tilt}}=4$, and $C_{\rm {bd}}=0$.
The circles represent $\Gamma$ calculated from a pore on the flat membrane at $N=2048$.
The solid line represents  $\Gamma$ calculated from the striped membrane at $N=512$: 
$\Gamma \sigma/k_{\rm B}T = 9.08$ $\pm 0.06$.
}
\end{figure}

\begin{figure}
\includegraphics{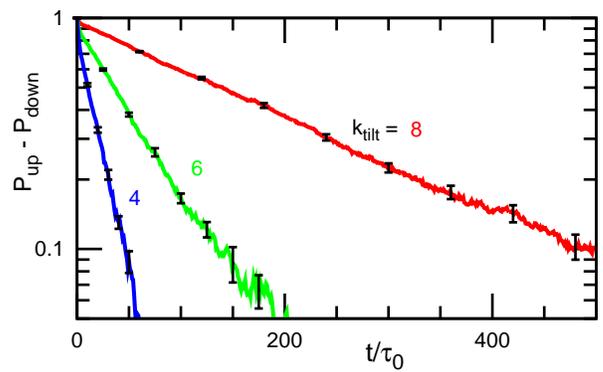}
\caption{\label{fig:flop}
(Color online)
Time development of the probability difference $P_{\rm {up}}-P_{\rm {down}}$ of 
the molecules in the upper and lower monolayers
at $N=512$, $\rho^*=14$, $\varepsilon=2$,
$k_{\rm {bend}}=0$, and $C_{\rm {bd}}=0$.
At the initial states ($t=0$), $P_{\rm {up}}=1$ and $P_{\rm {down}}=0$.
The error bars are displayed at several data points.
}
\end{figure}

The surface tension $\gamma$ is given by \cite{rowl82,alle87}
\begin{equation}
\gamma= \langle P_{zz} - (P_{xx} + P_{yy})/2\rangle L_z,
\label{eq:stpt}
\end{equation} 
with the diagonal components of the pressure tensor
\begin{equation}
\label{eq:pressure_tensor}
P_{\alpha\alpha} = (Nk_{\rm B}T - 
     \sum_{i} \alpha_{i}\frac{\partial U}{\partial {\alpha}_{i}} )/V,
\end{equation} 
where $\alpha \in \{x,y,z\}$.
When the potential interaction crosses the periodic boundary,
the periodic image $ \alpha_{i}+ n L_{\alpha}$ nearest to the other interacting molecules is employed.
The intrinsic area $A$ of the membrane is larger than the projected area 
$A_{xy}$ in the $xy$-plane due to the membrane undulations.
We calculate $A$ from a $\sqrt{N/2} \times \sqrt{N/2}$ square mesh with 
$(x_{\rm {mh}}, y_{\rm {mh}})=(d_{\rm {mh}}i, d_{\rm {mh}}j)$.
The height $z_{\rm {mh}}$ of a mesh point is obtained from the weighted 
average of molecular position ${\bf r}^{\rm s}_i$ in the four neighbor cells, with
$z_{\rm {mh}}=  \sum_i z_i w_{\rm {mh}}(x_i,y_i)/(\sum_i w_{\rm {mh}}(x_i,y_i))$
and $w_{\rm {mh}}(x_i,y_i) = 
(1-|x_i-x_{\rm {mh}}|/d_{\rm {mh}})(1-|y_i-y_{\rm {mh}}|/d_{\rm {mh}})$.
Figure \ref{fig:ten} shows the area dependence on the surface tension $\gamma$.
The tension $\gamma$ exhibits a roughly linear increase with the molecular area at $\gamma \gtrsim 0$.
The compressed membrane with $\gamma<0$  buckles out of plane 
and has the larger intrinsic area $A$ than the projected area $A_{xy}$.
Similar $\gamma$ dependence and buckling are obtained in the simulations of 
other molecular models \cite{otte05,dese09} 
and meshless models \cite{nogu06,kohy09}.

The area $A_0$ of the tensionless membrane ($\gamma=0$) is obtained by the minimization of $\gamma$,
where the projected area is updated 
as $A_{xy}^{\rm {new}}=A_{xy} - b\bar{\gamma} \Delta t_{\gamma}$ every $\Delta t_{\gamma}$ interval,
where $\bar{\gamma}$ is the time average for $\Delta t_{\gamma}$.
We use $b \tau_0=0.00025$ to $0.005$ and $\Delta t_{\gamma}/\tau_0=50$ or $100$.
The area compression modulus 
$K_{\rm A}$ is defined as
\begin{eqnarray}
K_{\rm A}= A_{\rm 0}\partial \gamma/\partial A|_{A=A_{\rm 0}}.
\label{eq:ka}
\end{eqnarray} 
We calculate $K_{\rm A}$ from the slope of $a$-$\gamma$ lines
shown in Fig. \ref{fig:ten}(b).

The bending rigidity $\kappa$ is calculated from the spectra of undulation
modes $\langle |h(q)|^2\rangle$ of the planar membranes in Fourier
space \cite{safr94,goet99,lind00},
\begin{equation}
\langle |h(q)|^2\rangle=\frac{k_{\rm B}T}{\gamma q^2 + \kappa q^4} .
\label{eq:hq}
\end{equation}
Figure~\ref{fig:hq} clearly shows the $q^{-4}$ dependence of the 
tensionless membrane.
We calculate $|h(q)|^2$ from the raw data (the particle position ${\bf r}^{\rm s}_i$), as 
well as from the square mesh with the same mesh-points which were used
for the estimation of the intrinsic  area $A$.
Averaging over the mesh removes most of the effects of the molecular
protrusions.
The bending rigidity $\kappa$ is estimated from a fit of 
$1/\langle |h(q)|^2\rangle= (\kappa/k_{\rm B}T)(q^2)^2$ for $(q/\pi)^2<0.015$,
where the difference of two spectra is very small
(see the inset of Fig.~\ref{fig:hq}).

The line tension $\Gamma$  of the membrane edge
is calculated 
from the strip of the flat membrane as \cite{tolp04,reyn08}
\begin{equation}
\Gamma =  \langle P_{xx} - (P_{yy} + P_{zz})/2\rangle L_{y}L_{z}/2,
\label{eq:lt}
\end{equation} 
since $\Gamma$ is the energy per unit length of the membrane edge, and
the length of the membrane edge  is $2L_{x}$.
Since the striped membrane is tensionless, 
$\langle P_{yy}\rangle=\langle P_{zz} \rangle \simeq 0$ for solvent-free simulations.
The tension $\Gamma$ and its error bar are estimated 
from the average and standard deviations for $L_{x}/\sigma=14$, $15$, $16$, and $18$.
Alternatively, $\Gamma$ can be also calculated from a circular pore on the flat membrane \cite{tolp04,nogu06}.
In this case, $\Gamma$ is balanced with the surface tension $\gamma$ as $\Gamma = \gamma r_{\rm {p0}}$.
Since  one has to estimate the pore radius $r_{\rm {p0}}$, this method gives larger statistical errors as shown in Fig. \ref{fig:line_r}.
Therefore, we used the membrane strip for the calculation of $\Gamma$.

\begin{figure}
\includegraphics[width=8.cm]{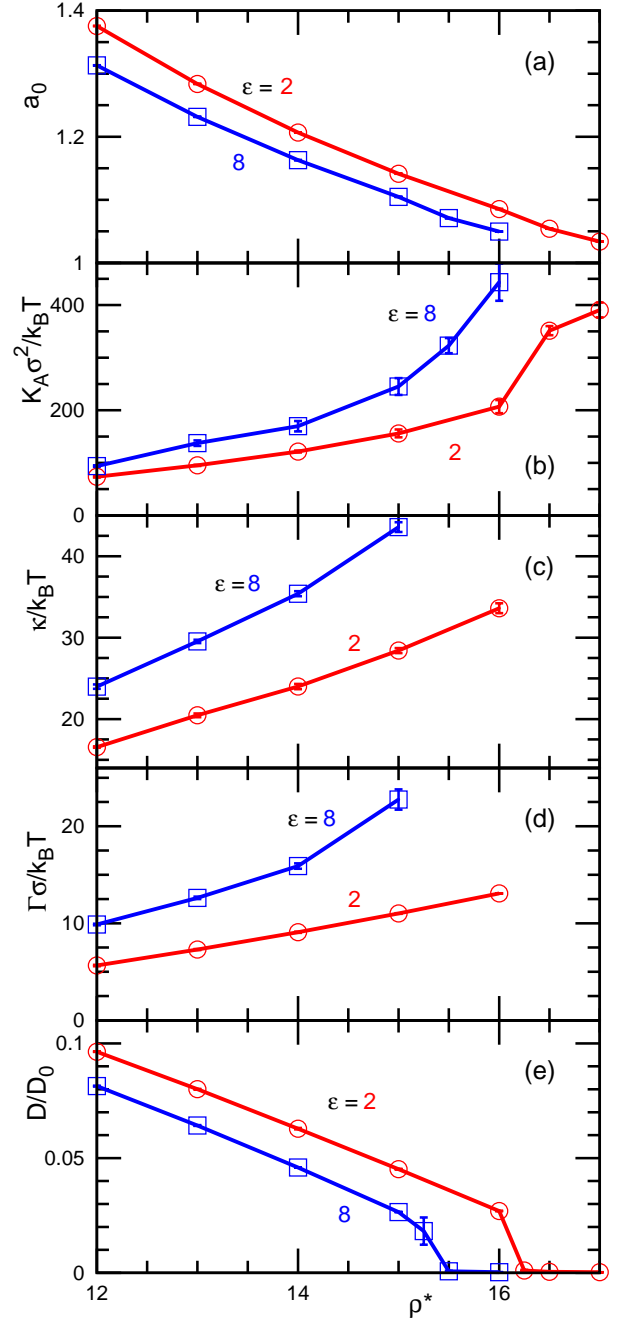}
\caption{\label{fig:mem_p}
(Color online)
Parameter $\rho^*$ dependence of (a) the intrinsic area $a_0=2A_0/N\sigma^2$ 
per molecule, (b) area compression modulus $K_{\rm A}$,
(c) bending rigidity $\kappa$, 
(d) line tension $\Gamma$, and
(e) diffusion coefficient $D$ for the tensionless membrane 
at $k_{\rm {tilt}}=4$, $k_{\rm {bend}}=4$, and $C_{\rm {bd}}=0$. 
The circles and squares represent data for $\varepsilon=2$ and $8$, respectively.
}
\end{figure}

\begin{figure}
\includegraphics{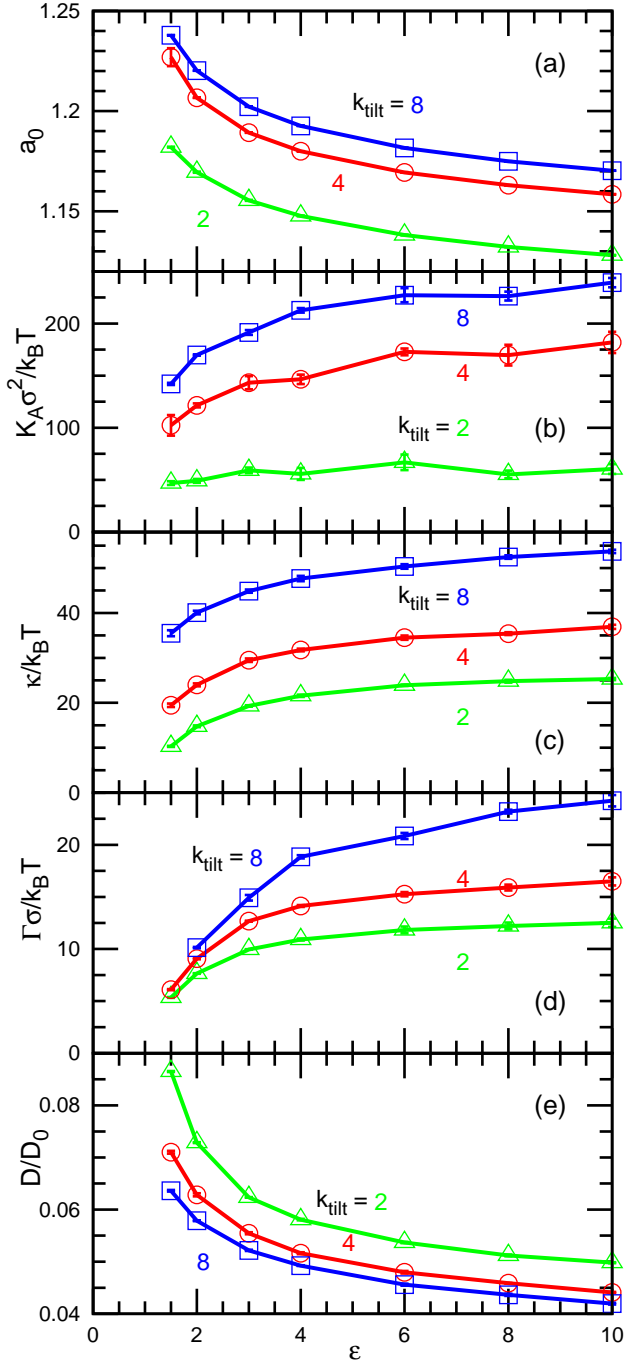}
\caption{\label{fig:mem_e}
(Color online)
Parameter $\varepsilon$ dependence of (a) $a_0=2A_0/N\sigma^2$,
 (b) $K_{\rm A}$, (c) $\kappa$, (d) $\Gamma$, and
(e) $D$ for $\rho^*=14$, $k_{\rm {bend}}=k_{\rm {tilt}}$, and $C_{\rm {bd}}=0$.
The triangles, circles, and squares represent data for $k_{\rm {tilt}}=2$, $4$, and $8$, respectively.
}
\end{figure}

In bilayer membranes,
molecules can move laterally on a monolayer and transversely between upper and lower monolayers.
The lateral diffusion coefficient $D$ of the molecules is calculated from the diffusion 
of the molecular projections in the $xy$ plane;
$D= \langle (x_i(t)-x_i(0))^2 + (y_i(t)-y_i(0))^2\rangle/4t$.
In the fluid phase, the molecules exhibit a fast diffusion rate $D/D_0 = 0.05$ to $0.1$.

The relaxation time of the transverse motion (flip-flop) between the upper and lower monolayers
 is measured from the relaxation of the labeled molecules \cite{korn71,nogu01a}.
The differential equation of the probability $P_{\rm {up}}(t)$ ($P_{\rm {down}}(t)$) of the molecules, 
which belong to the upper (lower) monolayer, is given by 
$dP_{\rm {up}}/dt= -k_{\rm u}P_{\rm {up}} + k_{\rm d}P_{\rm {down}}$,
where $P_{\rm {up}}+P_{\rm {down}}=1$.
For planar membranes, $k_{\rm u}=k_{\rm d}$, 
and $P_{\rm {up}}=P_{\rm {down}}=1/2$ at $t \to \infty$.
When the molecules in the upper monolayer are initially labeled ($P_{\rm {up}}(0)=1$),
the probability decays as
\begin{equation}
P_{\rm {up}}(t) - P_{\rm {down}}(t)= \exp[-(k_{\rm u}+k_{\rm d})t ]
\label{eq:ff}
\end{equation} 
with the half lifetime $\tau_{\rm {ff}}=\ln(2)/(k_{\rm u}+k_{\rm d})$.
Figure \ref{fig:flop} shows that $P_{\rm {up}}(t) - P_{\rm {down}}(t)$ indeed 
follows the exponential decay in our simulations.
Either the $z$ component of the position ${\bf r}^{\rm e}_i$ or orientation ${\bf u}_i$
can be used to detect the nomolayer, to which a molecule belongs.
Both ${\bf r}^{\rm e}_i$ and ${\bf u}_i$ give the same probability distribution 
(entirely same for most of the parameters),
since both the quantities have a clear minimum between two peaks (see Fig. \ref{fig:hisze}).

\begin{figure}
\includegraphics{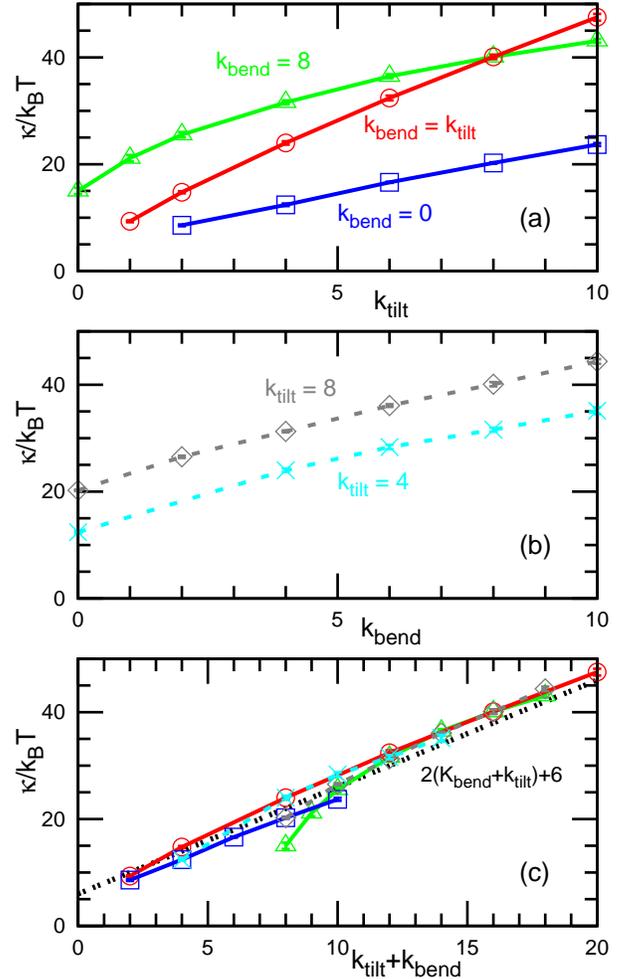}
\caption{\label{fig:kappa}
(Color online)
Bending rigidity $\kappa$ dependence on (a) $k_{\rm {tilt}}$, (b) $k_{\rm {bend}}$, 
and (c) $k_{\rm {tilt}}+k_{\rm {bend}}$ for $\rho^*=14$, $\varepsilon=2$, and $C_{\rm {bd}}=0$.
The solid lines with squares, circles, and triangles represent data 
for $k_{\rm {bend}}=0$, $k_{\rm {bend}}=k_{\rm {tilt}}$, and $k_{\rm {bend}}=8$, respectively.
The dashed lines with crosses and diamonds represent data for $k_{\rm {tilt}}=4$ and $8$, respectively.
}
\end{figure}

\begin{figure}
\includegraphics{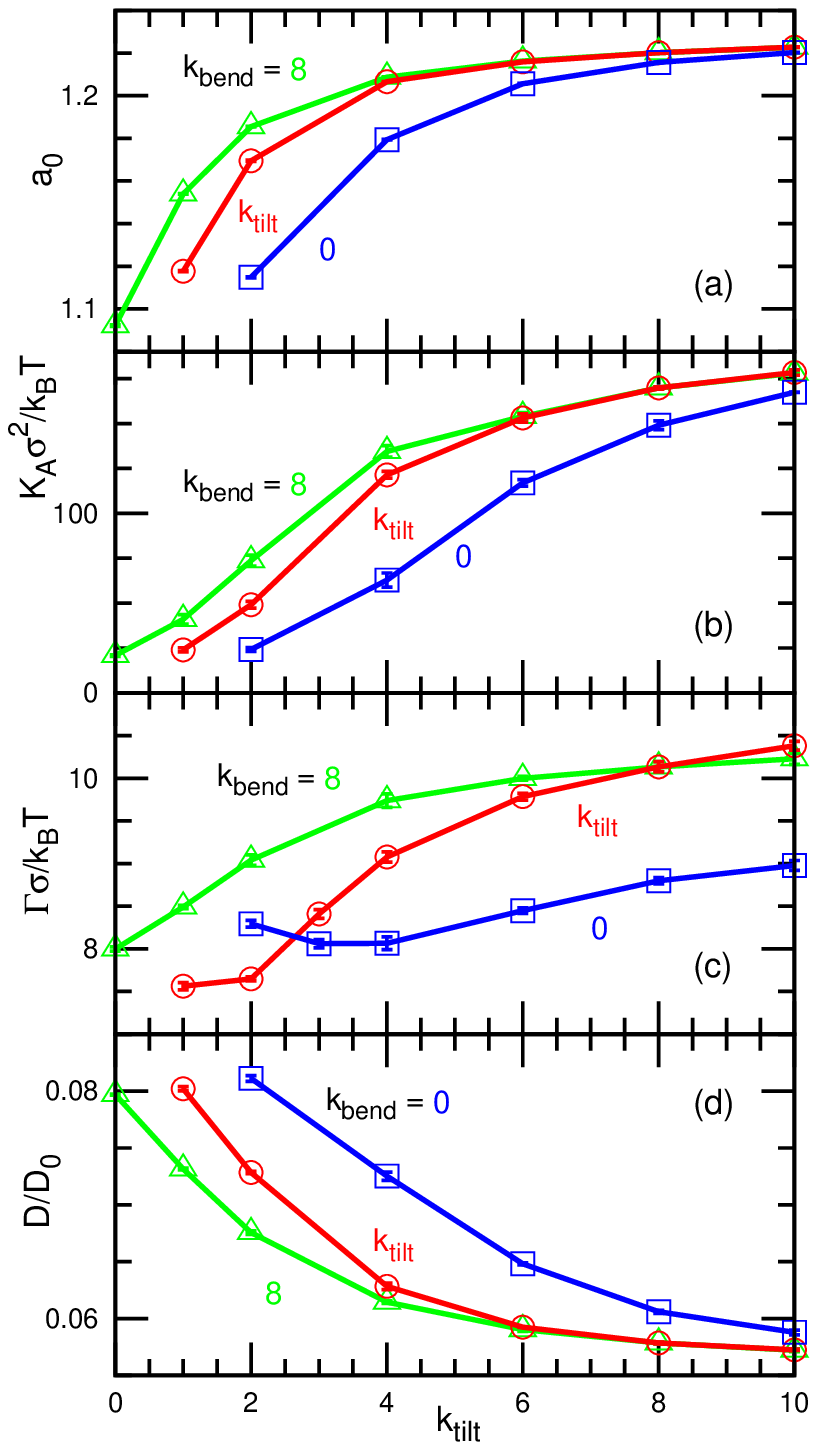}
\caption{\label{fig:mem_t}
Parameter $k_{\rm {tilt}}$ dependence of (a) $a_0=2A_0/N\sigma^2$,
 (b) $K_{\rm A}$, (c) $\Gamma$, and
(d) $D$ for $\rho^*=14$, $\varepsilon=2$, and $C_{\rm {bd}}=0$.
The squares, circles, and triangles represent data 
for $k_{\rm {bend}}=0$, $k_{\rm {bend}}=k_{\rm {tilt}}$, and $k_{\rm {bend}}=8$, respectively.
}
\end{figure}

\subsection{Parameter dependence of membrane properties}
\label{sec:paradep}

\begin{figure}
\includegraphics{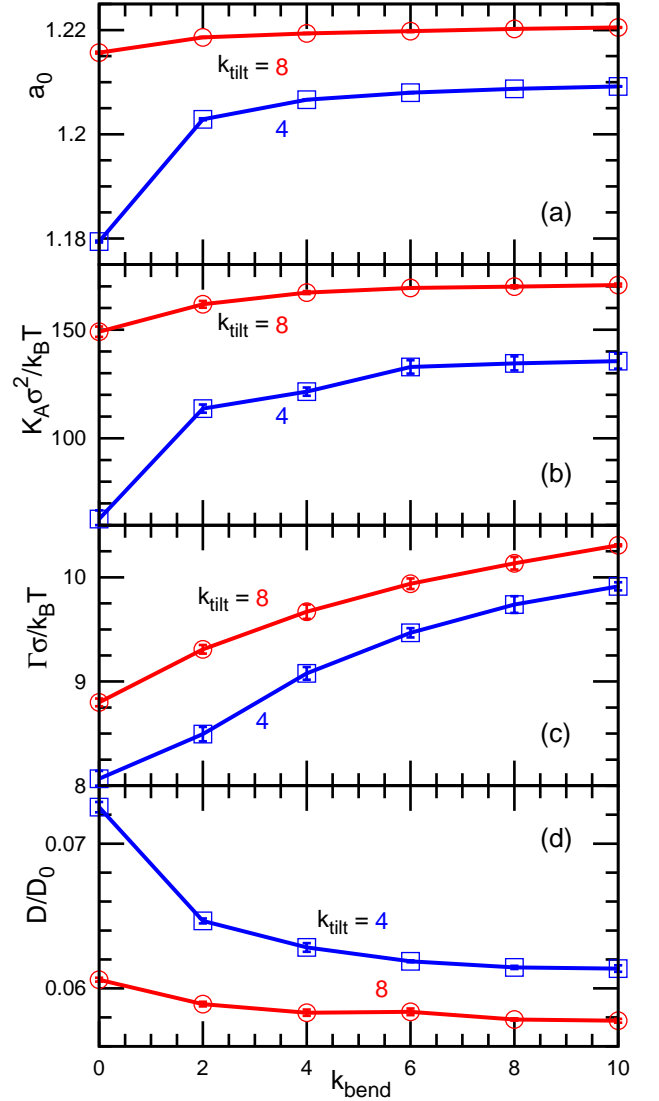}
\caption{\label{fig:mem_j}
(Color online)
Parameter $k_{\rm {bend}}$ dependence of (a) $a_0=2A_0/N\sigma^2$,
 (b) $K_{\rm A}$, (c) $\Gamma$, and
(d) $D$ for $\rho^*=14$, $\varepsilon=2$, and $C_{\rm {bd}}=0$.
The squares and circles represent data for $k_{\rm {tilt}}=4$ and $8$, respectively.
}
\end{figure}

\begin{figure}
\includegraphics{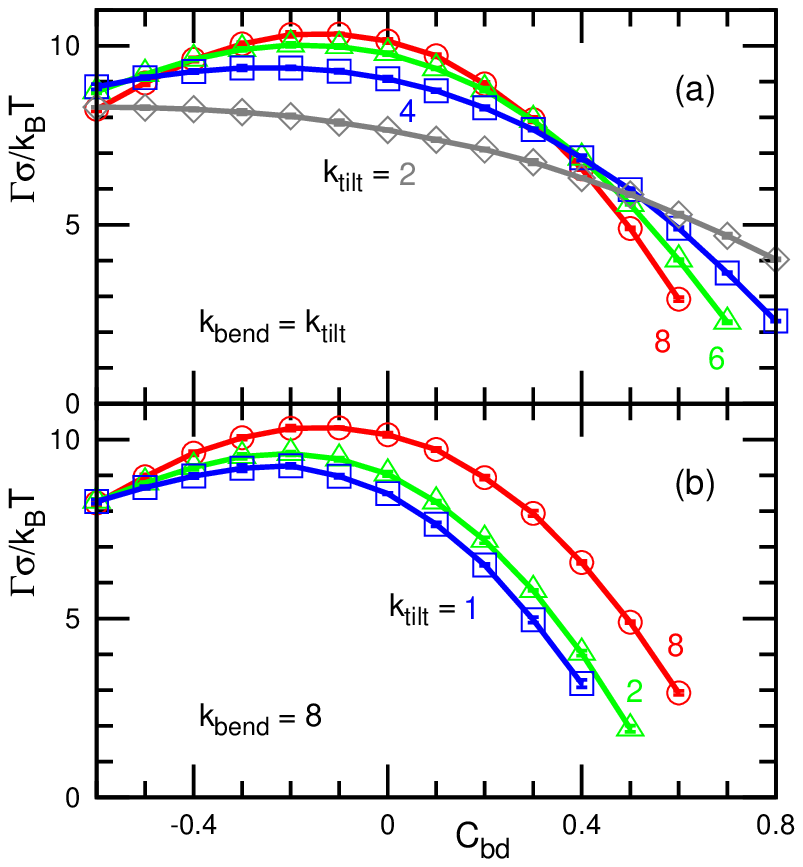}
\caption{\label{fig:line_c}
(Color online)
Line tension $\Gamma$ dependence on $C_{\rm {bd}}$ for (a) $k_{\rm {bend}}=k_{\rm {tilt}}$ and (b) $k_{\rm {bend}}=8$
at $\rho^*=14$ and $\varepsilon=2$.
(a) The diamonds, squares, triangles, and circles represent data for $k_{\rm {tilt}}=2, 4, 6$, and $8$, respectively.
(b) The squares, triangles, and circles represent data for $k_{\rm {tilt}}=1, 2$, and $8$, respectively.
}
\end{figure}

\begin{figure}
\includegraphics{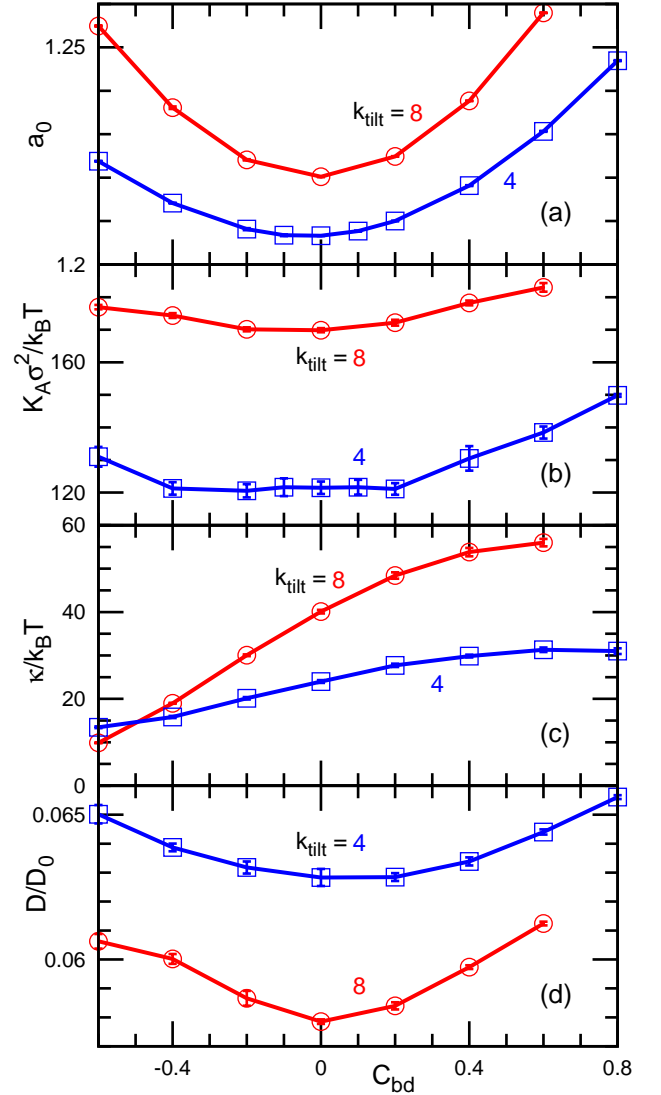}
\caption{\label{fig:mem_c}
(Color online)
Parameter $C_{\rm {bd}}$ dependence of (a) $a_0=2A_0/N\sigma^2$,
 (b) $K_{\rm A}$, (c) $\kappa$, and
(d) $D$ for $\rho^*=14$, $\varepsilon=2$, and $k_{\rm {bend}}=k_{\rm {tilt}}$.
The squares and circles represent data for $k_{\rm {tilt}}=4$ and $8$, respectively.
}
\end{figure}

\begin{figure}
\includegraphics[width=8.1cm]{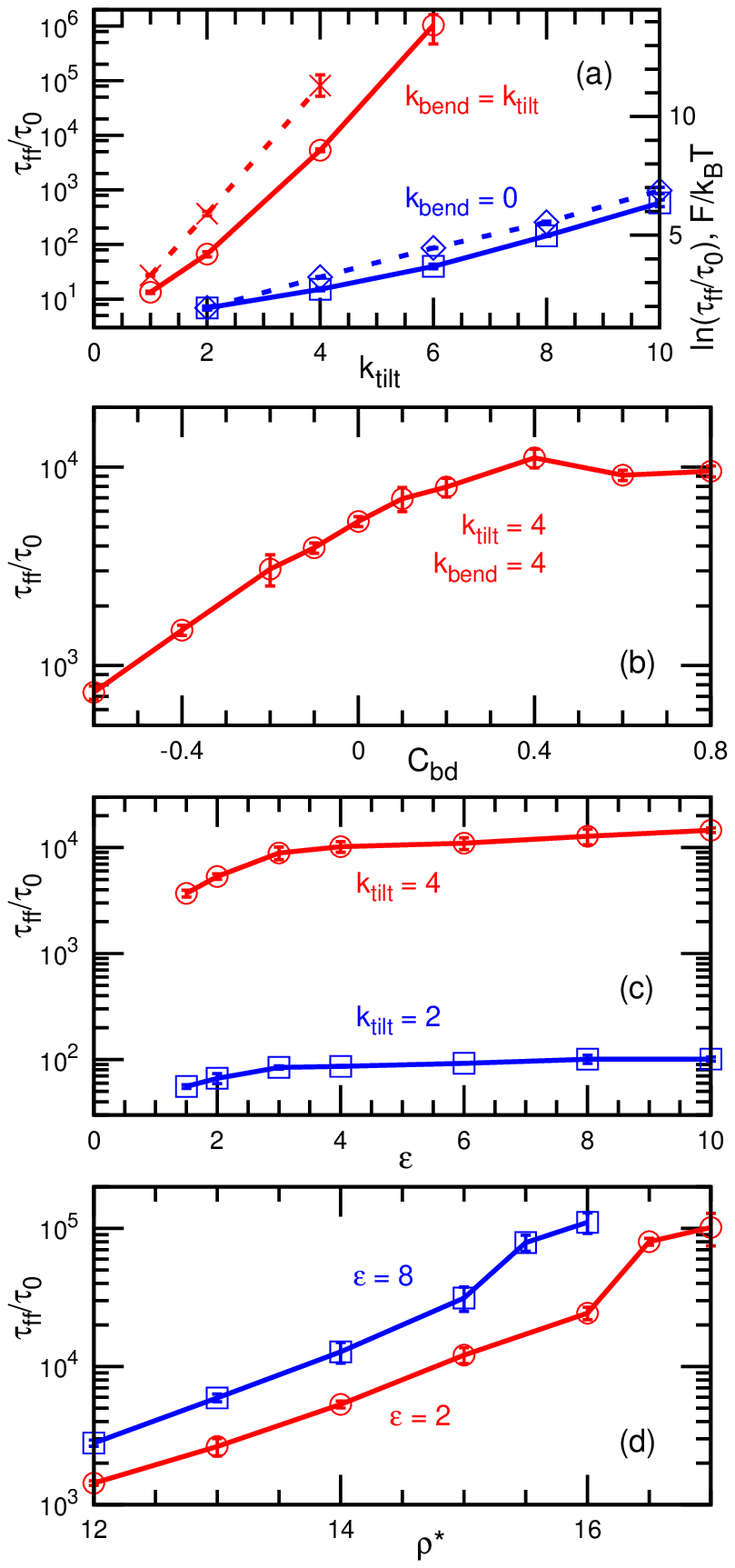}
\caption{\label{fig:ff}
(Color online)
Half lifetime $\tau_{\rm {ff}}$ of flip-flop motion.
(a) Dependence on $k_{\rm {tilt}}$ at $\rho^*=14$, $\varepsilon=2$, and $C_{\rm {bd}}=0$.
The dashed lines represent the free-energy barrier $F_{\rm {ff}}/k_{\rm B}T$
estimated by the orientation distribution shown in Fig. \ref{fig:hisze}(b).
(b) Dependence on $C_{\rm {bd}}$ at $\rho^*=14$ and $\varepsilon=2$.
(c) Dependence on $\varepsilon$ at  $\rho^*=14$, $k_{\rm {bend}}=k_{\rm {tilt}}$,  and $C_{\rm {bd}}=0$.
(d) Dependence on $\rho^*$ at $k_{\rm {bend}}=k_{\rm {tilt}}=4$ and $C_{\rm {bd}}=0$.
}
\end{figure}

Figures \ref{fig:mem_p}--\ref{fig:ff} show the parameter dependence of the properties
of the tensionless membrane.
The bilayer membrane is formed in the fluid phase over broad ranges of the parameters
 due to the multibody attractive potential.
A gel phase is obtained only at a large value of $\rho^*$ in Eq. (\ref{eq:U_att}).
The fluid-gel transition occurs at $\rho^* = 16$ and  $\rho^* = 15$ for $\varepsilon=2$ and $8$, 
respectively [see jumps of $D$ in  Fig. \ref{fig:mem_p}(e)].
As $\rho^*$ decreases, the lateral diffusion and flip-flop motion become faster,
and the membrane elasticities ($K_{\rm A}$, $\kappa$, $\Gamma$) decrease [see Figs. \ref{fig:mem_p} and \ref{fig:ff}(d)].
The  intrinsic area $a_0=2A_0/N\sigma^2$ per molecule ($N/2$ molecules in one of monolayers)
decreases with increasing $\rho^*$ and approaches 
the closest-packing area $\sqrt{3}\sigma^2/2 \simeq 0.87 \sigma^2$ in the gel phase.
In this paper, we focus on the fluid membrane and set $\rho^*=14$, hereafter.

The dependence on 
the strength of attraction $\varepsilon$ in Eq. (\ref{eq:U_all}) is shown in Fig. \ref{fig:mem_e}.
It has a tendency similar to $\rho^*$ dependence.
The line tension $\Gamma$ can be varied by $\varepsilon$.
At $\varepsilon=1$,
$\Gamma$ is close to $k_{\rm B}T/\sigma$ and
the bilayer membrane is accompanied by free molecules (gas)
with the average density of the gas $\sim 0.001/\sigma^3$.
The molecules depart from the bilayer membrane and return.
At $\varepsilon=0.75$ with $k_{\rm {bend}}=k_{\rm {tilt}}=4$ and $C_{\rm {bd}}=0$,
 the bilayer membrane breaks and small micelles 
are formed ($\simeq 60$ molecules per micelle for $N/V=0.08/\sigma^3$).
On the other hand, no free molecules are seen at $\varepsilon \geq 2$. 
The critical micelle concentration (CMC) of lipids 
is very low, and their chemical potential difference in solution
and in membrane is typically more than $10k_{\rm B}T$ per lipid \cite{tanf80}.
Thus, the number of lipid molecules on vesicles is conserved in  typical experiments.
In order to keep the number of molecules on membrane constant during simulations,
a sufficiently low CMC is required.
We mainly use $\varepsilon = 2$ and $8$,
where the fluid membranes without free molecules are obtained.

The bending rigidity $\kappa$ of the bilayer membrane can be controlled by $k_{\rm {tilt}}$ or $k_{\rm {bend}}$.
 in Eq. (\ref{eq:U_all}).
Figures \ref{fig:kappa}(a) and (b) show the linear dependence of $\kappa$ 
on $k_{\rm {tilt}}$ and $k_{\rm {bend}}$, respectively.
When they are plotted together for $k_{\rm {tilt}}+ k_{\rm {bend}}$,
all lines are roughly overlapped on a line $\kappa/k_{\rm B}T=2(k_{\rm {tilt}}+ k_{\rm {bend}})+6$ [see Fig. \ref{fig:kappa}(c)].
A large deviation from the line is seen only at one data point at $k_{\rm {tilt}}=0$ and $k_{\rm {bend}}=8$ 
(the leftmost triangle), where the bilayer structure is metastable.
The bending rigidity $\kappa$ is also weakly dependent on $\varepsilon$.
The $\kappa$-$\varepsilon$ curve maintains its shape and shifts upward with increasing $k_{\rm {tilt}}$ 
as shown in Fig. \ref{fig:mem_e}(c).
Thus, it is expressed by $\kappa/k_{\rm B}T=2(k_{\rm {tilt}}+ k_{\rm {bend}})+b_\varepsilon(\varepsilon)$,
where $b_\varepsilon(\varepsilon)$ is an increasing function as
$b_\varepsilon(2)=6$, $b_\varepsilon(4)=14$, and $b_\varepsilon(8)=18$.
The area compression modulus $K_{\rm A}$ increases with increasing $k_{\rm {tilt}}$,
while  $K_{\rm A}$ shows only slight dependence on $k_{\rm {bend}}$ for large $k_{\rm {tilt}}$  
(see Figs. \ref{fig:mem_t} and \ref{fig:mem_j}).
The other membrane properties $a$, $\Gamma$, and $D$ show
weak dependence on  $k_{\rm {tilt}}$ and $k_{\rm {bend}}$.
Thus, $\kappa$ can be varied by $k_{\rm {tilt}}$ or $k_{\rm {bend}}$ 
without a large variation in $\Gamma$.
The modulus $K_{\rm A}$ can be varied by $k_{\rm {tilt}}$.

\begin{figure}
\includegraphics{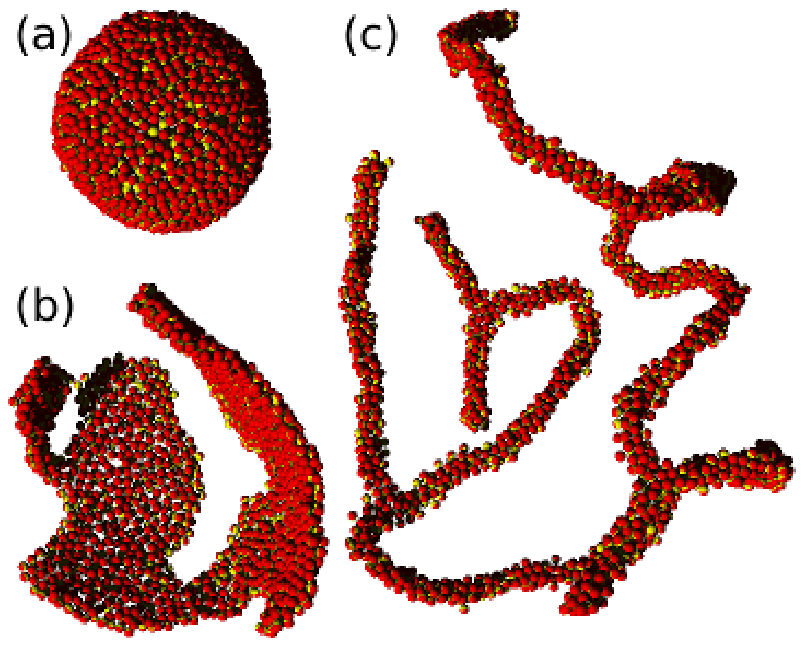}
\caption{\label{fig:rup_snap}
(Color online)
Sequential snapshots of vesicle rupture
at $N=2000$, $\varepsilon=2$, $\rho^*=14$, $k_{\rm {bend}}=8$, $k_{\rm {tilt}}=8$, and $C_{\rm {bd}}=0.85$.
(a) $t/\tau_0=500$. (b) $t/\tau_0=640$. (c) $t/\tau_0=3000$.
}
\end{figure}

The bending elastisity generated by the orientation-dependent potentials can be
derived from the Helfrich theory for monolayer membranes.
When the orientation vectors ${\bf u}_{i}$ are equal to the normal
 vectors of the monolayers without tilt deformation,
the bending and tilt energies are written by
\begin{eqnarray}
U_{\rm {cv}} &=& \int dA\  \frac{\kappa'_{\rm {bend}}}{2}[
(C_1-C'_0)^2 + (C_2-C'_0)^2] \nonumber \\ \label{eq:cv0} 
& & \hspace{0.9cm} + \frac{\kappa'_{\rm {tilt}}}{2}[(C_1^2 +C_2^2)] \\ \nonumber
 &=&  \int dA\  \frac{\kappa'_{\rm {bend}}+\kappa'_{\rm {tilt}}}{2}[
(C_1+C_2-C_0)^2   \\ \label{eq:cv1} 
& & \hspace{0.9cm} - (\kappa'_{\rm {bend}}+\kappa'_{\rm {tilt}})C_1 C_2 + U_0
\end{eqnarray}
in the continuum limit, where
$C_1$ and $C_2$ are two principal curvatures of
the monolayer.
The first and second terms in Eq. (\ref{eq:cv0}) are
the contributions of the bending and tilt potentials, respectively.
The spontaneous curvature of the bending potential is given by $C'_0= C_{\rm {bd}}/\bar{r}_{\rm {nb}}$. \cite{nogu03}
The nearest-neighbor distance $\bar{r}_{\rm {nb}} \simeq 1.05\sigma$ is obtained from the radial distribution function.
By assuming the hexagonal packing of the molecules in the monolayers,
the monolayer bending rigidities generated by  the bending and tilt potentials are estimated as 
$\kappa'_{\rm {bend}}/k_{\rm B}T= \sqrt{3} k_{\rm {bend}} w_{\rm {cv}}(\bar{r}_{\rm {nb}})$ and
$\kappa'_{\rm {tilt}}/k_{\rm B}T= \sqrt{3} k_{\rm {tilt}} w_{\rm {cv}}(\bar{r}_{\rm {nb}})/2$, respectively.
The bending rigidity of the monolayer is given by the sum of these  $\kappa_{\rm {mono}}=\kappa'_{\rm {bend}}+\kappa'_{\rm {tilt}}$.
For the monolayer membrane, Eq. (\ref{eq:cv1}) gives the saddle-splay modulus  $\bar{\kappa}_{\rm {mono}}= -\kappa_{\rm {mono}}$
and the spontaneous curvature $C_0= \{\kappa'_{\rm {bend}}/(\kappa'_{\rm {bend}}+\kappa'_{\rm {tilt}})\}C_{\rm {bd}}/\bar{r}_{\rm {nb}}$ 
with $U_0 =(\kappa'_{\rm {bend}}+\kappa'_{\rm {tilt}})(1/2+\kappa'_{\rm {tilt}}/\kappa'_{\rm {bend}})C_0^2$.
Thus, the bending rigidity $\kappa$  of the bilayer is estimated as 
$\kappa=2\kappa_{\rm {mono}} \simeq (2.1 k_{\rm {bend}} + 1.1 k_{\rm {tilt}})k_{\rm B}T$ from $w_{\rm {cv}}(1.05\sigma) = 0.61$.
This estimation of $\kappa$ supports the linear dependence of the obtained simulation results on $k_{\rm {bend}}$ and $k_{\rm {tilt}}$.
The prefactor ($\simeq 2$) of $k_{\rm {bend}}$ gives the quantitative agreement, whereas
the prefactor of $k_{\rm {tilt}}$ is half of the numerical estimation.
In the simulation, the thermal fluctuations induce molecular protrusion and tilt.
These tilt fluctuations likely change the prefactor of $k_{\rm {tilt}}$ to twice its value
($\kappa'_{\rm {tilt}}/\kappa'_{\rm {bend}}=k_{\rm {tilt}}/k_{\rm {bend}}$).
The attractive potential also adds a small bending resistance, $b_\varepsilon(\varepsilon)$.

The flip-flop time $\tau_{\rm {ff}}$ shows exponential dependence on the parameters
of the tilt and bending potentials, while it has weak dependence on $\varepsilon$ (see Fig. \ref{fig:ff}).
The free-energy barrier between two monolayers would be the main factor to determine $\tau_{\rm {ff}}$.
It can be roughly estimated from the probability distribution of the molecular orientation
 as  $F_{\rm {ff}}=k_{\rm B}T(\ln[P(u_z=1)] - \ln[P(u_z=0)])$ [see Fig. \ref{fig:hisze}(b)].
The dashed lines in  Fig. \ref{fig:ff}(a) represent  $F_{\rm {ff}}$ calculated by this method.
It gives very good agreements with $\ln(\tau_{\rm {ff}}/\tau_0)$, and thus,
the flip-flop time is written by $\tau_{\rm {ff}}/\tau_0 = b_{\rm {ff}}\exp(F_{\rm {ff}}/k_{\rm B}T)$ 
with $b_{\rm {ff}} \simeq 1$.
The barrier $F_{\rm {ff}}$ is linear with
the orientation-dependent potentials, while it is almost independent of the attractive potentials.
The flip-flop rate can be tuned by $k_{\rm {tilt}}$ or  $k_{\rm {bend}}$.
In typical experimental conditions, the flip-flop motion of phospholipids
 is very slow, and $\tau_{\rm {ff}}$ is several hours or days \cite{korn71}.
At a sufficiently high $k_{\rm {tilt}}$ or  $k_{\rm {bend}}$,
no flip-flop occurs in typical simulation time scales.
This result agrees with the experimental observations.
On the other hand, at small $k_{\rm {tilt}}$ and $k_{\rm {bend}}$,
the molecule shows  very fast flip-flop,
which is advantageous to equilibrate the membrane system quickly during simulations.
In the present model, one can choose slow or fast flip-flop condition to match one's simulation purpose.

The spontaneous curvature of the monolayer is varied by the parameter $C_{\rm {bd}}$.
According to the above continuous theory,
$C_0= \{k_{\rm {bend}}/\sigma(k_{\rm {bend}}+k_{\rm {tilt}})\}C_{\rm {bd}}$.
At high spontaneous curvature $C_0 \sim 1/\sigma$, a worm-like micelle is stabilized.
As $C_{\rm {bd}}$ increases, the line tension $\Gamma$ decreases, and 
 the bilayer structure becomes unstable at $\Gamma \sim k_{\rm B}T/\sigma$.
The $C_{\rm {bd}}$-$\Gamma$ curves in Fig. \ref{fig:line_c} have a parabolic shape with a maximum 
at $C_{\rm {bd}} \sim -0.1$.
The bending rigidity $\kappa$ and the flip-flop time $\tau_{\rm {ff}}$
increase with $C_{\rm {bd}}$, while $a_0$, $K_{\rm A}$, and $D$ do not depend significantly on $C_{\rm {bd}}$
[see Figs. \ref{fig:mem_c} and \ref{fig:ff}(b)]. 
At negative $C_0$, the molecules easily stay at the middle of the bilayer structure (low $F_{\rm {ff}}$),
and the less clear bilayer would generate lower $\kappa$.

Membrane rupture is observed when $C_{\rm {bd}}$ is increased.
Figure \ref{fig:rup_snap} shows the rupture process of a vesicle at $C_{\rm {bd}}=0.85$.
The initial vesicle is  metastable and maintains its shape for $500\tau_0$.
Then, a pore opens and grows to cracks [see Fig. \ref{fig:rup_snap}(b)].
Finally, a branched worm-like micelle is formed [see Fig. \ref{fig:rup_snap}(c)].
Thus, the line tension and the stable structure can be varied by $C_{\rm {bd}}$.

\section{Summary}
\label{sec:sum}

We have proposed a simple molecular model of bilayer membranes.
The molecule has five degrees of freedom, three translational degrees and two orientational degrees. 
Since this molecular model consists of one spherical particle for the excluded volume,
it is smaller than previous molecular models.
Thus, this model is more efficient for large-scale simulations.
Despite the extreme simplification, this model 
reproduces many aspects of lipid bilayer membranes.

In the present model, the properties of the fluid membranes can be controlled in broad ranges
including metastable bilayer membranes.
The bending rigidity $\kappa$
is linearly dependent on $k_{\rm {tilt}}$ and $k_{\rm {bend}}$.
The line tension $\Gamma$ of the membrane edge can be varied by 
$\varepsilon$ and $C_{\rm {bd}}$.
The membrane has a wide range of fluid phase, and
the fluid-gel transition point can be controlled by $\rho^*$.
The area compression modulus $K_{\rm A}$ can be varied by  $k_{\rm {tilt}}$.
The flip-flop time $\tau_{\rm {ff}}$ can be varied by $k_{\rm {tilt}}$ and $k_{\rm {bend}}$.

The corresponding time and length scales can be mapped by
the membrane thickness  $5$nm  and 
the lateral diffusion coefficient $\sim 10^{-8}$cm$^2$/s for phospholipids \cite{wu77}.
Thus, the unit length and times are estimated as $\sigma=2$ nm
and $\tau_{\rm 0} \sim 0.1 \mu$s, respectively.

Our model is suitable to study the details of topological-change processes of 
the membrane, such as molecular self-assembly, pore formation, membrane fusion, and membrane fission.
Although meshless membrane models can also be used,
they cannot treat detailed structures such as the fusion intermediates \cite{jahn02,cher08}.
Recently, fusion dynamics have been investigated by molecular simulations \cite{nogu01b,nogu02a,nogu02c,muel02,li05,shil05,smei06,knec07}.
However, the condition to determine the fusion pathways has not be clarified so far.
The ability to vary the membrane properties in wide ranges would be an advantage of this model
for quantitative investigation of the membrane fusion pathways.
On the other hand, this model 
 is not suitable for application to phenomena, in which the stretching of hydrophobic chains  or
atomistic details play an important role,
e.g., in the interactions of membrane proteins via hydrophibic mismatch, where
molecular stretching is not negligible \cite{phil09,four99,deme08,west09}.

In this paper, we used Brownian dynamics
but one can also use the Monte Carlo method and molecular dynamics with another thermostat.
When the solvent-free molecular model is combined with a particle-based hydrodynamics method, 
multi-particle collision dynamics (MPC) \cite{kapr08,gg:gomp09a},
the hydrodynamic interaction can be taken into account as demonstrated for a meshless membrane model \cite{nogu06a}.
Thus, the present coarse-grained molecular model is efficient for large-scale dynamics of a biomembrane
with or without hydrodynamic interactions and is applicable to many kinds of phenomena.

\begin{acknowledgments}
This work is supported by KAKENHI (21740308) from
the Ministry of Education, Culture, Sports, Science, and Technology of Japan.
\end{acknowledgments}


\begin{thebibliography}{57}
\expandafter\ifx\csname natexlab\endcsname\relax\def\natexlab#1{#1}\fi
\expandafter\ifx\csname bibnamefont\endcsname\relax
  \def\bibnamefont#1{#1}\fi
\expandafter\ifx\csname bibfnamefont\endcsname\relax
  \def\bibfnamefont#1{#1}\fi
\expandafter\ifx\csname citenamefont\endcsname\relax
  \def\citenamefont#1{#1}\fi
\expandafter\ifx\csname url\endcsname\relax
  \def\url#1{\texttt{#1}}\fi
\expandafter\ifx\csname urlprefix\endcsname\relax\def\urlprefix{URL }\fi
\providecommand{\bibinfo}[2]{#2}
\providecommand{\eprint}[2][]{\url{#2}}

\bibitem[{\citenamefont{Safran}(1994)}]{safr94}
\bibinfo{author}{\bibfnamefont{S.~A.} \bibnamefont{Safran}},
  \emph{\bibinfo{title}{Statistical Thermodynamics of Surfaces, Interfaces, and
  Membranes}} (\bibinfo{publisher}{Addison-Wesley}, \bibinfo{address}{Reading,
  MA}, \bibinfo{year}{1994}).

\bibitem[{\citenamefont{Lipowsky and Sackmann}(1995)}]{lipo95}
\bibinfo{editor}{\bibfnamefont{R.}~\bibnamefont{Lipowsky}} \bibnamefont{and}
  \bibinfo{editor}{\bibfnamefont{E.}~\bibnamefont{Sackmann}}, eds.,
  \emph{\bibinfo{title}{Structure and Dynamics of Membranes}}
  (\bibinfo{publisher}{Elsevier Science}, \bibinfo{address}{Amsterdam},
  \bibinfo{year}{1995}).

\bibitem[{\citenamefont{Seifert}(1997)}]{seif97}
\bibinfo{author}{\bibfnamefont{U.}~\bibnamefont{Seifert}},
  \bibinfo{journal}{Adv.\ Phys.} \textbf{\bibinfo{volume}{46}},
  \bibinfo{pages}{13} (\bibinfo{year}{1997}).

\bibitem[{\citenamefont{Gompper and Kroll}(2004)}]{gg:gomp04c}
\bibinfo{author}{\bibfnamefont{G.}~\bibnamefont{Gompper}} \bibnamefont{and}
  \bibinfo{author}{\bibfnamefont{D.~M.} \bibnamefont{Kroll}}, in
  \emph{\bibinfo{booktitle}{Statistical Mechanics of Membranes and Surfaces}},
  edited by \bibinfo{editor}{\bibfnamefont{D.~R.} \bibnamefont{Nelson}},
  \bibinfo{editor}{\bibfnamefont{T.}~\bibnamefont{Piran}}, \bibnamefont{and}
  \bibinfo{editor}{\bibfnamefont{S.}~\bibnamefont{Weinberg}}
  (\bibinfo{publisher}{World Scientific}, \bibinfo{address}{Singapore},
  \bibinfo{year}{2004}), \bibinfo{edition}{2nd} ed.

\bibitem[{\citenamefont{Gompper and Kroll}(1997)}]{gg:gomp97f}
\bibinfo{author}{\bibfnamefont{G.}~\bibnamefont{Gompper}} \bibnamefont{and}
  \bibinfo{author}{\bibfnamefont{D.~M.} \bibnamefont{Kroll}},
  \bibinfo{journal}{J. Phys. Condens. Matter} \textbf{\bibinfo{volume}{9}},
  \bibinfo{pages}{8795} (\bibinfo{year}{1997}).

\bibitem[{\citenamefont{Drouffe et~al.}(1991)\citenamefont{Drouffe, Maggs, and
  Leibler}}]{drou91}
\bibinfo{author}{\bibfnamefont{J.~M.} \bibnamefont{Drouffe}},
  \bibinfo{author}{\bibfnamefont{A.~C.} \bibnamefont{Maggs}}, \bibnamefont{and}
  \bibinfo{author}{\bibfnamefont{S.}~\bibnamefont{Leibler}},
  \bibinfo{journal}{Science} \textbf{\bibinfo{volume}{254}},
  \bibinfo{pages}{1353} (\bibinfo{year}{1991}).

\bibitem[{\citenamefont{Noguchi and Gompper}(2006{\natexlab{a}})}]{nogu06}
\bibinfo{author}{\bibfnamefont{H.}~\bibnamefont{Noguchi}} \bibnamefont{and}
  \bibinfo{author}{\bibfnamefont{G.}~\bibnamefont{Gompper}},
  \bibinfo{journal}{Phys.\ Rev.\ E} \textbf{\bibinfo{volume}{73}},
  \bibinfo{pages}{021903} (\bibinfo{year}{2006}{\natexlab{a}}).

\bibitem[{\citenamefont{Noguchi and Gompper}(2006{\natexlab{b}})}]{nogu06a}
\bibinfo{author}{\bibfnamefont{H.}~\bibnamefont{Noguchi}} \bibnamefont{and}
  \bibinfo{author}{\bibfnamefont{G.}~\bibnamefont{Gompper}},
  \bibinfo{journal}{J.\ Chem.\ Phys.} \textbf{\bibinfo{volume}{125}},
  \bibinfo{pages}{164908} (\bibinfo{year}{2006}{\natexlab{b}}).

\bibitem[{\citenamefont{{Del P{\'o}polo} and Ballone}(2008)}]{popo08}
\bibinfo{author}{\bibfnamefont{M.~G.} \bibnamefont{{Del P{\'o}polo}}}
  \bibnamefont{and} \bibinfo{author}{\bibfnamefont{P.}~\bibnamefont{Ballone}},
  \bibinfo{journal}{J.\ Chem.\ Phys.} \textbf{\bibinfo{volume}{128}},
  \bibinfo{pages}{024705} (\bibinfo{year}{2008}).

\bibitem[{\citenamefont{Kohyama}(2009)}]{kohy09}
\bibinfo{author}{\bibfnamefont{T.}~\bibnamefont{Kohyama}},
  \bibinfo{journal}{Physica A} \textbf{\bibinfo{volume}{388}},
  \bibinfo{pages}{3334} (\bibinfo{year}{2009}).

\bibitem[{\citenamefont{Liu et~al.}(2009)\citenamefont{Liu, Li, and
  Zhang}}]{liu09}
\bibinfo{author}{\bibfnamefont{P.}~\bibnamefont{Liu}},
  \bibinfo{author}{\bibfnamefont{J.}~\bibnamefont{Li}}, \bibnamefont{and}
  \bibinfo{author}{\bibfnamefont{Y.~W.} \bibnamefont{Zhang}},
  \bibinfo{journal}{Appl.\ Phys.\ Lett.} \textbf{\bibinfo{volume}{95}},
  \bibinfo{pages}{143104} (\bibinfo{year}{2009}).

\bibitem[{\citenamefont{F{\"u}chslin et~al.}(2009)\citenamefont{F{\"u}chslin,
  Maeke, and McCaskill}}]{fuch09}
\bibinfo{author}{\bibfnamefont{R.~M.} \bibnamefont{F{\"u}chslin}},
  \bibinfo{author}{\bibfnamefont{T.}~\bibnamefont{Maeke}}, \bibnamefont{and}
  \bibinfo{author}{\bibfnamefont{J.~S.} \bibnamefont{McCaskill}},
  \bibinfo{journal}{Eur.\ Phys.\ J.\ E} \textbf{\bibinfo{volume}{29}},
  \bibinfo{pages}{431} (\bibinfo{year}{2009}).

\bibitem[{\citenamefont{Yuan et~al.}(2010)\citenamefont{Yuan, Huang, Li,
  Lykotrafitis, and Zhang}}]{yuan10}
\bibinfo{author}{\bibfnamefont{H.}~\bibnamefont{Yuan}},
  \bibinfo{author}{\bibfnamefont{C.}~\bibnamefont{Huang}},
  \bibinfo{author}{\bibfnamefont{J.}~\bibnamefont{Li}},
  \bibinfo{author}{\bibfnamefont{G.}~\bibnamefont{Lykotrafitis}},
  \bibnamefont{and} \bibinfo{author}{\bibfnamefont{S.}~\bibnamefont{Zhang}},
  \bibinfo{journal}{Phys.\ Rev.\ E} \textbf{\bibinfo{volume}{82}},
  \bibinfo{pages}{011905} (\bibinfo{year}{2010}).

\bibitem[{\citenamefont{Jahn and Grubm{\"u}ller}(2002)}]{jahn02}
\bibinfo{author}{\bibfnamefont{R.}~\bibnamefont{Jahn}} \bibnamefont{and}
  \bibinfo{author}{\bibfnamefont{H.}~\bibnamefont{Grubm{\"u}ller}},
  \bibinfo{journal}{Curr.\ Opin.\ Cell\ Biol.} \textbf{\bibinfo{volume}{14}},
  \bibinfo{pages}{488} (\bibinfo{year}{2002}).

\bibitem[{\citenamefont{Chernomordik and Kozlov}(2008)}]{cher08}
\bibinfo{author}{\bibfnamefont{L.~V.} \bibnamefont{Chernomordik}}
  \bibnamefont{and} \bibinfo{author}{\bibfnamefont{M.~M.}
  \bibnamefont{Kozlov}}, \bibinfo{journal}{Nat.\ Struct.\ Mol.\ Biol.}
  \textbf{\bibinfo{volume}{15}}, \bibinfo{pages}{675} (\bibinfo{year}{2008}).

\bibitem[{\citenamefont{M{\"u}ller et~al.}(2006)\citenamefont{M{\"u}ller,
  Katsov, and Schick}}]{muel06}
\bibinfo{author}{\bibfnamefont{M.}~\bibnamefont{M{\"u}ller}},
  \bibinfo{author}{\bibfnamefont{K.}~\bibnamefont{Katsov}}, \bibnamefont{and}
  \bibinfo{author}{\bibfnamefont{M.}~\bibnamefont{Schick}},
  \bibinfo{journal}{Phys.\ Rep.} \textbf{\bibinfo{volume}{434}},
  \bibinfo{pages}{113} (\bibinfo{year}{2006}).

\bibitem[{\citenamefont{Venturoli et~al.}(2006)\citenamefont{Venturoli,
  Sperotto, Kranenburg, and Smit}}]{vent06}
\bibinfo{author}{\bibfnamefont{M.}~\bibnamefont{Venturoli}},
  \bibinfo{author}{\bibfnamefont{M.~M.} \bibnamefont{Sperotto}},
  \bibinfo{author}{\bibfnamefont{M.}~\bibnamefont{Kranenburg}},
  \bibnamefont{and} \bibinfo{author}{\bibfnamefont{B.}~\bibnamefont{Smit}},
  \bibinfo{journal}{Phys.\ Rep.} \textbf{\bibinfo{volume}{437}},
  \bibinfo{pages}{1} (\bibinfo{year}{2006}).

\bibitem[{\citenamefont{Klein and Shinoda}(2008)}]{klie08}
\bibinfo{author}{\bibfnamefont{M.~L.} \bibnamefont{Klein}} \bibnamefont{and}
  \bibinfo{author}{\bibfnamefont{W.}~\bibnamefont{Shinoda}},
  \bibinfo{journal}{Science} \textbf{\bibinfo{volume}{321}},
  \bibinfo{pages}{798} (\bibinfo{year}{2008}).

\bibitem[{\citenamefont{Noguchi}(2009)}]{nogu09}
\bibinfo{author}{\bibfnamefont{H.}~\bibnamefont{Noguchi}},
  \bibinfo{journal}{J.\ Phys.\ Soc.\ Jpn.} \textbf{\bibinfo{volume}{78}},
  \bibinfo{pages}{041007} (\bibinfo{year}{2009}).

\bibitem[{\citenamefont{Marrink et~al.}(2009)\citenamefont{Marrink, de\ Vries,
  and Tieleman}}]{marr09}
\bibinfo{author}{\bibfnamefont{S.~J.} \bibnamefont{Marrink}},
  \bibinfo{author}{\bibfnamefont{A.~H.} \bibnamefont{de\ Vries}},
  \bibnamefont{and} \bibinfo{author}{\bibfnamefont{D.~P.}
  \bibnamefont{Tieleman}}, \bibinfo{journal}{Biochim.\ Biophys.\ Acta}
  \textbf{\bibinfo{volume}{1788}}, \bibinfo{pages}{149} (\bibinfo{year}{2009}).

\bibitem[{\citenamefont{Marrink et~al.}(2004)\citenamefont{Marrink, de~Vries,
  and Mark}}]{marr04}
\bibinfo{author}{\bibfnamefont{S.~J.} \bibnamefont{Marrink}},
  \bibinfo{author}{\bibfnamefont{A.~H.} \bibnamefont{de~Vries}},
  \bibnamefont{and} \bibinfo{author}{\bibfnamefont{A.~E.} \bibnamefont{Mark}},
  \bibinfo{journal}{J.\ Phys.\ Chem.\ B} \textbf{\bibinfo{volume}{108}},
  \bibinfo{pages}{750} (\bibinfo{year}{2004}).

\bibitem[{\citenamefont{Izvekov and Voth}(2005)}]{izve05}
\bibinfo{author}{\bibfnamefont{S.}~\bibnamefont{Izvekov}} \bibnamefont{and}
  \bibinfo{author}{\bibfnamefont{G.~A.} \bibnamefont{Voth}},
  \bibinfo{journal}{J.\ Phys.\ Chem.\ B} \textbf{\bibinfo{volume}{109}},
  \bibinfo{pages}{2469} (\bibinfo{year}{2005}).

\bibitem[{\citenamefont{Arkhipov et~al.}(2008)\citenamefont{Arkhipov, Yin, and
  Schulten}}]{arkh08}
\bibinfo{author}{\bibfnamefont{A.}~\bibnamefont{Arkhipov}},
  \bibinfo{author}{\bibfnamefont{Y.}~\bibnamefont{Yin}}, \bibnamefont{and}
  \bibinfo{author}{\bibfnamefont{K.}~\bibnamefont{Schulten}},
  \bibinfo{journal}{Biophys.\ J.} \textbf{\bibinfo{volume}{95}},
  \bibinfo{pages}{2806} (\bibinfo{year}{2008}).

\bibitem[{\citenamefont{Shinoda et~al.}(2008)\citenamefont{Shinoda, DeVane, and
  Klein}}]{shin08}
\bibinfo{author}{\bibfnamefont{W.}~\bibnamefont{Shinoda}},
  \bibinfo{author}{\bibfnamefont{R.}~\bibnamefont{DeVane}}, \bibnamefont{and}
  \bibinfo{author}{\bibfnamefont{M.~L.} \bibnamefont{Klein}},
  \bibinfo{journal}{Soft Matter} \textbf{\bibinfo{volume}{4}},
  \bibinfo{pages}{2454} (\bibinfo{year}{2008}).

\bibitem[{\citenamefont{Wang and Deserno}(2010)}]{wang10}
\bibinfo{author}{\bibfnamefont{Z.~J.} \bibnamefont{Wang}} \bibnamefont{and}
  \bibinfo{author}{\bibfnamefont{M.}~\bibnamefont{Deserno}},
  \bibinfo{journal}{J.\ Phys.\ Chem.\ B} \textbf{\bibinfo{volume}{114}},
  \bibinfo{pages}{11207} (\bibinfo{year}{2010}).

\bibitem[{\citenamefont{Noguchi and Takasu}(2001{\natexlab{a}})}]{nogu01a}
\bibinfo{author}{\bibfnamefont{H.}~\bibnamefont{Noguchi}} \bibnamefont{and}
  \bibinfo{author}{\bibfnamefont{M.}~\bibnamefont{Takasu}},
  \bibinfo{journal}{Phys.\ Rev.\ E} \textbf{\bibinfo{volume}{64}},
  \bibinfo{pages}{041913} (\bibinfo{year}{2001}{\natexlab{a}}).

\bibitem[{\citenamefont{Noguchi}(2003)}]{nogu03}
\bibinfo{author}{\bibfnamefont{H.}~\bibnamefont{Noguchi}},
  \bibinfo{journal}{Phys.\ Rev.\ E} \textbf{\bibinfo{volume}{67}},
  \bibinfo{pages}{041901} (\bibinfo{year}{2003}).

\bibitem[{\citenamefont{Brannigan et~al.}(2006)\citenamefont{Brannigan, Lin,
  and Brown}}]{bran06}
\bibinfo{author}{\bibfnamefont{G.}~\bibnamefont{Brannigan}},
  \bibinfo{author}{\bibfnamefont{L.~C.~L.} \bibnamefont{Lin}},
  \bibnamefont{and} \bibinfo{author}{\bibfnamefont{F.~L.~H.}
  \bibnamefont{Brown}}, \bibinfo{journal}{Eur. Biophys. J.}
  \textbf{\bibinfo{volume}{35}}, \bibinfo{pages}{104} (\bibinfo{year}{2006}).

\bibitem[{\citenamefont{Deserno}(2009)}]{dese09}
\bibinfo{author}{\bibfnamefont{M.}~\bibnamefont{Deserno}},
  \bibinfo{journal}{Macromol.\ Rapid.\ Commun.} \textbf{\bibinfo{volume}{30}},
  \bibinfo{pages}{752} (\bibinfo{year}{2009}).

\bibitem[{\citenamefont{Farago and Gr{\o}nbech-Jensen}(2009)}]{fara09}
\bibinfo{author}{\bibfnamefont{O.}~\bibnamefont{Farago}} \bibnamefont{and}
  \bibinfo{author}{\bibfnamefont{N.}~\bibnamefont{Gr{\o}nbech-Jensen}},
  \bibinfo{journal}{J.\ Am.\ Chem.\ Soc.} \textbf{\bibinfo{volume}{131}},
  \bibinfo{pages}{2875} (\bibinfo{year}{2009}).

\bibitem[{\citenamefont{Takada et~al.}(1999)\citenamefont{Takada,
  Luthey-Schulten, and Wolynes}}]{taka99}
\bibinfo{author}{\bibfnamefont{S.}~\bibnamefont{Takada}},
  \bibinfo{author}{\bibfnamefont{Z.}~\bibnamefont{Luthey-Schulten}},
  \bibnamefont{and} \bibinfo{author}{\bibfnamefont{P.~G.}
  \bibnamefont{Wolynes}}, \bibinfo{journal}{J.\ Chem.\ Phys.}
  \textbf{\bibinfo{volume}{110}}, \bibinfo{pages}{11616}
  (\bibinfo{year}{1999}).

\bibitem[{\citenamefont{Hamm and Kozlov}(1998)}]{hamm98}
\bibinfo{author}{\bibfnamefont{M.}~\bibnamefont{Hamm}} \bibnamefont{and}
  \bibinfo{author}{\bibfnamefont{M.~M.} \bibnamefont{Kozlov}},
  \bibinfo{journal}{Eur.\ Phys.\ J.\ B} \textbf{\bibinfo{volume}{6}},
  \bibinfo{pages}{519} (\bibinfo{year}{1998}).

\bibitem[{\citenamefont{Hamm and Kozlov}(2000)}]{hamm00}
\bibinfo{author}{\bibfnamefont{M.}~\bibnamefont{Hamm}} \bibnamefont{and}
  \bibinfo{author}{\bibfnamefont{M.~M.} \bibnamefont{Kozlov}},
  \bibinfo{journal}{Eur.\ Phys.\ J.\ E} \textbf{\bibinfo{volume}{3}},
  \bibinfo{pages}{323} (\bibinfo{year}{2000}).

\bibitem[{\citenamefont{Allen and Tildesley}(1987)}]{alle87}
\bibinfo{author}{\bibfnamefont{M.~P.} \bibnamefont{Allen}} \bibnamefont{and}
  \bibinfo{author}{\bibfnamefont{D.~J.} \bibnamefont{Tildesley}},
  \emph{\bibinfo{title}{Computer simulation of liquids}}
  (\bibinfo{publisher}{Clarendon Press}, \bibinfo{address}{Oxford},
  \bibinfo{year}{1987}).

\bibitem[{\citenamefont{Rowlinson and Widom}(1982)}]{rowl82}
\bibinfo{author}{\bibfnamefont{J.~S.} \bibnamefont{Rowlinson}}
  \bibnamefont{and} \bibinfo{author}{\bibfnamefont{B.}~\bibnamefont{Widom}},
  \emph{\bibinfo{title}{Molecular Theory of Capillarity}}
  (\bibinfo{publisher}{Clarendon}, \bibinfo{address}{Oxford},
  \bibinfo{year}{1982}).

\bibitem[{\citenamefont{den Otter}(2005)}]{otte05}
\bibinfo{author}{\bibfnamefont{W.~K.} \bibnamefont{den Otter}},
  \bibinfo{journal}{J.\ Chem.\ Phys.} \textbf{\bibinfo{volume}{123}},
  \bibinfo{pages}{214906} (\bibinfo{year}{2005}).

\bibitem[{\citenamefont{Goetz et~al.}(1999)\citenamefont{Goetz, Gompper, and
  Lipowsky}}]{goet99}
\bibinfo{author}{\bibfnamefont{R.}~\bibnamefont{Goetz}},
  \bibinfo{author}{\bibfnamefont{G.}~\bibnamefont{Gompper}}, \bibnamefont{and}
  \bibinfo{author}{\bibfnamefont{R.}~\bibnamefont{Lipowsky}},
  \bibinfo{journal}{Phys.\ Rev.\ Lett.} \textbf{\bibinfo{volume}{82}},
  \bibinfo{pages}{221} (\bibinfo{year}{1999}).

\bibitem[{\citenamefont{Lindahl and Edholm}(2000)}]{lind00}
\bibinfo{author}{\bibfnamefont{E.}~\bibnamefont{Lindahl}} \bibnamefont{and}
  \bibinfo{author}{\bibfnamefont{O.}~\bibnamefont{Edholm}},
  \bibinfo{journal}{Biophys.\ J.} \textbf{\bibinfo{volume}{79}},
  \bibinfo{pages}{426} (\bibinfo{year}{2000}).

\bibitem[{\citenamefont{Tolpekina et~al.}(2004)\citenamefont{Tolpekina, den
  Otter, and Briels}}]{tolp04}
\bibinfo{author}{\bibfnamefont{T.~V.} \bibnamefont{Tolpekina}},
  \bibinfo{author}{\bibfnamefont{W.~K.} \bibnamefont{den Otter}},
  \bibnamefont{and} \bibinfo{author}{\bibfnamefont{W.~J.}
  \bibnamefont{Briels}}, \bibinfo{journal}{J.\ Chem.\ Phys.}
  \textbf{\bibinfo{volume}{121}}, \bibinfo{pages}{8014} (\bibinfo{year}{2004}).

\bibitem[{\citenamefont{Reynwar and Deserno}(2008)}]{reyn08}
\bibinfo{author}{\bibfnamefont{B.~J.} \bibnamefont{Reynwar}} \bibnamefont{and}
  \bibinfo{author}{\bibfnamefont{M.}~\bibnamefont{Deserno}},
  \bibinfo{journal}{Bioinerphases} \textbf{\bibinfo{volume}{3}},
  \bibinfo{pages}{FA117} (\bibinfo{year}{2008}).

\bibitem[{\citenamefont{Kornberg and McConnell}(1971)}]{korn71}
\bibinfo{author}{\bibfnamefont{R.~D.} \bibnamefont{Kornberg}} \bibnamefont{and}
  \bibinfo{author}{\bibfnamefont{H.~M.} \bibnamefont{McConnell}},
  \bibinfo{journal}{Biochemistry} \textbf{\bibinfo{volume}{10}},
  \bibinfo{pages}{1111} (\bibinfo{year}{1971}).

\bibitem[{\citenamefont{Tanford}(1980)}]{tanf80}
\bibinfo{author}{\bibfnamefont{C.}~\bibnamefont{Tanford}},
  \emph{\bibinfo{title}{The hydrophobic effect: formation of micelles and
  biological membranes}} (\bibinfo{publisher}{Wiley}, \bibinfo{address}{New
  York}, \bibinfo{year}{1980}).

\bibitem[{\citenamefont{Wu et~al.}(1977)\citenamefont{Wu, Jacobson, and
  Papahadjopoulos}}]{wu77}
\bibinfo{author}{\bibfnamefont{E.~S.} \bibnamefont{Wu}},
  \bibinfo{author}{\bibfnamefont{K.}~\bibnamefont{Jacobson}}, \bibnamefont{and}
  \bibinfo{author}{\bibfnamefont{D.}~\bibnamefont{Papahadjopoulos}},
  \bibinfo{journal}{Biochemistry} \textbf{\bibinfo{volume}{16}},
  \bibinfo{pages}{3936} (\bibinfo{year}{1977}).

\bibitem[{\citenamefont{Noguchi and Takasu}(2001{\natexlab{b}})}]{nogu01b}
\bibinfo{author}{\bibfnamefont{H.}~\bibnamefont{Noguchi}} \bibnamefont{and}
  \bibinfo{author}{\bibfnamefont{M.}~\bibnamefont{Takasu}},
  \bibinfo{journal}{J.\ Chem.\ Phys.} \textbf{\bibinfo{volume}{115}},
  \bibinfo{pages}{9547} (\bibinfo{year}{2001}{\natexlab{b}}).

\bibitem[{\citenamefont{Noguchi and Takasu}(2002)}]{nogu02a}
\bibinfo{author}{\bibfnamefont{H.}~\bibnamefont{Noguchi}} \bibnamefont{and}
  \bibinfo{author}{\bibfnamefont{M.}~\bibnamefont{Takasu}},
  \bibinfo{journal}{Biophys.\ J.} \textbf{\bibinfo{volume}{83}},
  \bibinfo{pages}{299} (\bibinfo{year}{2002}).

\bibitem[{\citenamefont{Noguchi}(2002)}]{nogu02c}
\bibinfo{author}{\bibfnamefont{H.}~\bibnamefont{Noguchi}},
  \bibinfo{journal}{J.\ Chem.\ Phys.} \textbf{\bibinfo{volume}{117}},
  \bibinfo{pages}{8130} (\bibinfo{year}{2002}).

\bibitem[{\citenamefont{M{\"u}ller et~al.}(2002)\citenamefont{M{\"u}ller,
  Katsov, and Schick}}]{muel02}
\bibinfo{author}{\bibfnamefont{M.}~\bibnamefont{M{\"u}ller}},
  \bibinfo{author}{\bibfnamefont{K.}~\bibnamefont{Katsov}}, \bibnamefont{and}
  \bibinfo{author}{\bibfnamefont{M.}~\bibnamefont{Schick}},
  \bibinfo{journal}{J. Chem. Phys.} \textbf{\bibinfo{volume}{116}},
  \bibinfo{pages}{2342} (\bibinfo{year}{2002}).

\bibitem[{\citenamefont{Li and Liu}(2005)}]{li05}
\bibinfo{author}{\bibfnamefont{D.~W.} \bibnamefont{Li}} \bibnamefont{and}
  \bibinfo{author}{\bibfnamefont{X.~Y.} \bibnamefont{Liu}},
  \bibinfo{journal}{J.\ Chem.\ Phys.} \textbf{\bibinfo{volume}{122}},
  \bibinfo{pages}{174909} (\bibinfo{year}{2005}).

\bibitem[{\citenamefont{Shillcock and Lipowsky}(2005)}]{shil05}
\bibinfo{author}{\bibfnamefont{J.}~\bibnamefont{Shillcock}} \bibnamefont{and}
  \bibinfo{author}{\bibfnamefont{R.}~\bibnamefont{Lipowsky}},
  \bibinfo{journal}{Nat.\ Mater.} \textbf{\bibinfo{volume}{4}},
  \bibinfo{pages}{225} (\bibinfo{year}{2005}).

\bibitem[{\citenamefont{Smeijers et~al.}(2006)\citenamefont{Smeijers,
  Markvoort, Pieterse, and Hilbers}}]{smei06}
\bibinfo{author}{\bibfnamefont{A.~F.} \bibnamefont{Smeijers}},
  \bibinfo{author}{\bibfnamefont{A.~J.} \bibnamefont{Markvoort}},
  \bibinfo{author}{\bibfnamefont{K.}~\bibnamefont{Pieterse}}, \bibnamefont{and}
  \bibinfo{author}{\bibfnamefont{P.~A.~J.} \bibnamefont{Hilbers}},
  \bibinfo{journal}{J. Phys.\ Chem.\ B} \textbf{\bibinfo{volume}{110}},
  \bibinfo{pages}{13212} (\bibinfo{year}{2006}).

\bibitem[{\citenamefont{Knecht and Marrink}(2007)}]{knec07}
\bibinfo{author}{\bibfnamefont{V.}~\bibnamefont{Knecht}} \bibnamefont{and}
  \bibinfo{author}{\bibfnamefont{S.~J.} \bibnamefont{Marrink}},
  \bibinfo{journal}{Biophys.\ J.} \textbf{\bibinfo{volume}{92}},
  \bibinfo{pages}{4254} (\bibinfo{year}{2007}).

\bibitem[{\citenamefont{Phillips et~al.}(2009)\citenamefont{Phillips, Ursell,
  Wiggins, and Sens}}]{phil09}
\bibinfo{author}{\bibfnamefont{R.}~\bibnamefont{Phillips}},
  \bibinfo{author}{\bibfnamefont{T.}~\bibnamefont{Ursell}},
  \bibinfo{author}{\bibfnamefont{P.}~\bibnamefont{Wiggins}}, \bibnamefont{and}
  \bibinfo{author}{\bibfnamefont{P.}~\bibnamefont{Sens}},
  \bibinfo{journal}{Nature} \textbf{\bibinfo{volume}{459}},
  \bibinfo{pages}{379} (\bibinfo{year}{2009}).

\bibitem[{\citenamefont{Fournier}(1999)}]{four99}
\bibinfo{author}{\bibfnamefont{J.~B.} \bibnamefont{Fournier}},
  \bibinfo{journal}{Eur.\ Phys.\ J. B} \textbf{\bibinfo{volume}{11}},
  \bibinfo{pages}{261} (\bibinfo{year}{1999}).

\bibitem[{\citenamefont{{de Meyer} et~al.}(2008)\citenamefont{{de Meyer},
  Venturoli, and Smit}}]{deme08}
\bibinfo{author}{\bibfnamefont{F.~J.} \bibnamefont{{de Meyer}}},
  \bibinfo{author}{\bibfnamefont{M.}~\bibnamefont{Venturoli}},
  \bibnamefont{and} \bibinfo{author}{\bibfnamefont{B.}~\bibnamefont{Smit}},
  \bibinfo{journal}{Biophys.\ J.} \textbf{\bibinfo{volume}{95}},
  \bibinfo{pages}{1851} (\bibinfo{year}{2008}).

\bibitem[{\citenamefont{West et~al.}(2009)\citenamefont{West, Brown, and
  Schmid}}]{west09}
\bibinfo{author}{\bibfnamefont{B.}~\bibnamefont{West}},
  \bibinfo{author}{\bibfnamefont{F.~L.~H.} \bibnamefont{Brown}},
  \bibnamefont{and} \bibinfo{author}{\bibfnamefont{F.}~\bibnamefont{Schmid}},
  \bibinfo{journal}{Biophys.\ J.} \textbf{\bibinfo{volume}{96}},
  \bibinfo{pages}{101} (\bibinfo{year}{2009}).

\bibitem[{\citenamefont{Kapral}(2008)}]{kapr08}
\bibinfo{author}{\bibfnamefont{R.}~\bibnamefont{Kapral}},
  \bibinfo{journal}{Adv. Chem. Phys.} \textbf{\bibinfo{volume}{140}},
  \bibinfo{pages}{89} (\bibinfo{year}{2008}).

\bibitem[{\citenamefont{Gompper et~al.}(2009)\citenamefont{Gompper, Ihle,
  Kroll, and Winkler}}]{gg:gomp09a}
\bibinfo{author}{\bibfnamefont{G.}~\bibnamefont{Gompper}},
  \bibinfo{author}{\bibfnamefont{T.}~\bibnamefont{Ihle}},
  \bibinfo{author}{\bibfnamefont{D.~M.} \bibnamefont{Kroll}}, \bibnamefont{and}
  \bibinfo{author}{\bibfnamefont{R.~G.} \bibnamefont{Winkler}},
  \bibinfo{journal}{Adv. Polym. Sci.} \textbf{\bibinfo{volume}{221}},
  \bibinfo{pages}{1} (\bibinfo{year}{2009}).

\end{thebibliography}

\end{document}